\documentclass[%
aps,
 amsmath,amssymb,
 reprint
]{revtex4-2}
\usepackage{graphicx}
\usepackage{dcolumn}
\usepackage[utf8]{inputenc}
\usepackage[T1]{fontenc}
\usepackage{mathptmx}
\usepackage{etoolbox}
\usepackage{dsfont}
\usepackage[colorlinks=true, citecolor=blue, linkcolor=blue, urlcolor=blue]{hyperref}

\renewcommand{\vec}[1]{{\bf #1}}

\newcommand{\tr}{\text{tr}}

\makeatletter
\def\@email#1#2{%
 \endgroup
 \patchcmd{\titleblock@produce}
  {\frontmatter@RRAPformat}
  {\frontmatter@RRAPformat{\produce@RRAP{*#1\href{mailto:#2}{#2}}}\frontmatter@RRAPformat}
  {}{}
}%
\makeatother

\begin{document}

\preprint{AIP/123-QED}
\title{Two-photon absorption cross sections of pulsed entangled beams}

\author{Frank Schlawin}
\affiliation 
{Max Planck Institute for the Structure and Dynamics of Matter, Luruper Chaussee 149, 22761 Hamburg, Germany}
\affiliation{University of Hamburg, Luruper Chaussee 149, Hamburg, Germany}
\affiliation{The Hamburg Centre for Ultrafast Imaging, Hamburg, Germany}
\email{frank.schlawin@mpsd.mpg.de}

\date{\today}

\begin{abstract}

Entangled two-photon absorption (ETPA) could form the basis of nonlinear quantum spectroscopy at very low photon fluxes, since, at sufficiently low photon fluxes, ETPA scales linearly with the photon flux. 
When different pairs start to overlap temporally, accidental coincidences are thought to give rise to a 'classical' quadratic scaling which dominates the signal at large photon fluxes and thus recovers a supposedly classical regime, where any quantum advantage is thought to be lost. 
Here we scrutinize this assumption and demonstrate that quantum-enhanced absorption cross sections can persist even to very large photon numbers. To this end, we use a minimal model for quantum light, which can interpolate continuously between the entangled pair and a high-photon-flux limit, to derive analytically ETPA cross sections and the intensity crossover regime. We investigate the interplay between spectral and spatial degrees of freedom, how linewidth broadening of the sample impacts the experimentally achievable enhancement.

\end{abstract}

\maketitle

\section{Introduction}

It was recognized already in the early days of quantum optics that the quantum statistics of a light field is intimately connected to the nonlinear optical response it generates in a material~\cite{Mollow, Shen67}. 
But it is only in recent years with the advent of high-flux quantum light sources that this nonlinear regime of quantum light-matter interactions has become experimentally accessible~\cite{Dayan05, Lee06, Spasibko17}. 
These experiments have since triggered a rapidly growing interest in the use of quantum light in nonlinear spectroscopy~\cite{Dorfman16, AccChemRes, Eshun2022, Szoke2020, Mukamel2020}. 
Current investigations range from the analysis of the photon statistics of light emitted by photosynthetic complexes~\cite{Siyuan17, Li2023, Yang2020, Dorfman2021, Carlos2020} to the improvement of the signal-to-noise ratio in stimulated Raman scattering~\cite{deAndrade2020, Casacio2021,Taylor2016}.

Arguably the most active field involves the use of entangled photons as spectroscopic tools~\cite{JPhysB2017}. 
The archetypical process, which is widely believed to encapsulate all the beneficial properties of entangled photons, is ETPA, where an entangled pair is absorbed by a quantum system via far off-resonant intermediate states. 
For such a process to occur, it is necessary for both photons to be localised at the position of the quantum system at the same time. This implies that the efficiency of this process may depend on the spatial and the spectral correlations of the entangled pair - in addition to the bunched nature of the entangled light field. 
In 1997 Fei et al. first condensed the interplay of these effects into the elegant formula for the ETPA cross section,~\cite{Fei1997}
\begin{align}\label{eq.ETPA-cross-section_Fei}
\sigma_e \simeq \frac{\delta_r}{A_e T_e}.
\end{align}
Here,  $\delta_r$ is the TPA cross section of randomly arriving photons. It is divided  by the so-called "entanglement area" $A_e$ and the "entanglement time" $T_e$, which quantify the abovementioned momentum correlations and spectral correlations, respectively. 
This result has since been generalized to multiphoton cross sections~\cite{Perina1998}. 
The above cross section gives rise to the detected ETPA rate,~\cite{Fei1997}
\begin{align}\label{eq.ETPA-rate_Fei}
R^{TPA} &= \sigma_e \phi + \delta_r \phi^2
\end{align}
where $\phi$ is the photon flux density. 
Since entangled photons always arrive in pairs, their ETPA rate scales linearly in the photon flux density~\cite{Gea89, Javainen90, Dayan07}. 
When the flux increases, such that different pairs overlap in time, random coincidences between uncorrelated photons are expected to give rise to the classical quadratic scaling of the second term. 

Despite the paramount importance of the two equations, they have never been studied systematically in experiments~\cite{CaracasNunez23}. 
Only one aspect - the linear scaling of the ETPA rate, which is arguably  by far their most important prediction, has been investigated intensely. 
It was first observed in atomic samples~\cite{Georgiades95, Dayan04, Dayan05}, and 
a tremendous amount of research has since focused on extending these results to molecular samples. 
Some experiments report very large ETPA signals in complex molecules~\cite{Lee06, Guzman10, Upton13,Villabona18, Varnavski2017, Eshun18, Varnavski2020, Villabona2020, Eshun2021, Burdick2021, Varnavski2023}, with the help of coincidence detection~\cite{Villabona17}, or with higher intensity squeezed states~\cite{Li20b}.
Only one series of experiments explicitly demonstrated the anticipated ETPA signal scaling after attenuation of the entangled photon flux~\cite{Tabakaev2021, Tabakaev2022}.
Moreover, a growing number of experiments failed to detect any clear ETPA signal at all~\cite{Parzuchowski2021,Landes2021, Lerch2021, Hickam2022, Corona-Aquino2022, Gaebler2023}. 
Beyond the scaling behaviour, the dependence on the spatial and temporal correlations, as predicted by Eq.~(\ref{eq.ETPA-cross-section_Fei}), was not studied to date.

Theoretical arguments were advanced for why weak ETPA signals should be expected in molecular samples~\cite{Raymer2020,Landes2021b,Raymer2021,Raymer2021b}, highlighting the interplay between molecular linewidth broadening and the achievable efficiency with the use of spectrally entangled photons.
Other theory works propose that entangled photons may couple efficiently to very narrow transitions~\cite{Kang2020, Giri2022}. 
Further investigations compared the relative strength of one-photon transitions which compete with the sought-after two-photon absorption~\cite{Nakanishi2009, Raymer2020, Drago2022, Raymer2022, Chen2022, Martinez2023}, highlighted the role of interference between excitation pathways~\cite{Wittkop2023}, the optimization of measurements~\cite{Mazurek19}, or considered the possibility of spectrally shaping the entangled two-photon wavefunction in order to optimise the ETPA efficiency~\cite{NJP2017, Oka2018, Carnio2021, Carnio2021b}.  Quantum metrology was used to analyze the achievable enhancement compared to classical laser spectroscopy~\cite{Carlos2021, Candeloro2021, Panahiyan2022, Panahiyan2023, Albarelli2023,Khan2023}. 

In this paper, we aim to further this discussion by analyzing molecular ETPA cross sections for pulsed, broadband entangled photons, taking into account both spatial and spectral degrees of freedom. We use a minimal model for entangled photons, which captures all the important physics, but allows us to obtain analytical results, to first derive Eq.~(\ref{eq.ETPA-cross-section_Fei}), and subsequently assess the validity of Eq.~(\ref{eq.ETPA-rate_Fei}).
In the former goal, we show that Eq.~(\ref{eq.ETPA-cross-section_Fei}) has to be amended by a penalty factor which depends on the ratio between the broadening of the molecular resonance and the bandwidth of the laser pulse generating the entangled beam. 
In the latter case, we show that Eq.~(\ref{eq.ETPA-rate_Fei}) is incorrect, as coherent effects can survive up to very large photon numbers. 
In the limit of a single spatial mode and a narrowband pump laser, our results largely agree with the analysis by Raymer and Landes in Ref.~\onlinecite{Raymer2022}. 
The paper is structured as follows: 
In Section~\ref{sec.model}, we discuss this model and highlight important properties. 
Section~\ref{sec.TPA} contains the perturbative calculations of the two terms in Eq.~(\ref{eq.ETPA-rate_Fei}). 
In Section~\ref{sec.two-photons}, we then restrict our attention to the two-photon limit, and derive Eq.~(\ref{eq.ETPA-cross-section_Fei}) from our microscopic model. 
Going beyond the two-photon limit, we discuss two limiting cases: in Section~\ref{sec.single-spatial-mode}, we consider the situation where a single spatial mode is sufficient to describe the light field, and in Section~\ref{sec.single-spectral-mode} we consider ETPA with only a single active spectral mode. 

\section{Light-matter interactions \& quantum light model}
\label{sec.model}

\subsection{The quantized electromagnetic field}

We want to describe the situation sketched in Fig.~\ref{fig.setup}(a), where a nonlinear crystal generates quantum light propagating along the $z$-direction with a small transverse momentum $\vec{q}$. A thin slab containing sample molecules is placed in the image plane of the optical system, where we set the origin of our coordinate system $z = 0$. 
In an ideal optical system, the positive frequency component of the electric field operator in the image plane can then be written as~\cite{Tsang2007}
\begin{align}
\vec{E}^{(+)} (\vec{\rho}, z=0, t) &= \frac{i}{(2\pi)^{3/2}} \sum_\sigma \int_0^\infty \!\!d\omega \; \int_{\mathds{R}^2} \!\!\!\! d^2 q  \left( \frac{\hbar \omega}{2 \epsilon_0} \right)^{1/2}  \notag \\
&\times\left(\frac{\omega}{c^2 k_z (\omega)}\right)^{1/2} \hat{a}_\sigma (\vec{q}, \omega) \vec{e}_{\vec{q},\sigma} e^{i (\vec{q}\cdot\vec{\rho} - \omega t)}, \label{eq.E(r,t)}
\end{align}
where $\hat{a}_\sigma (\vec{q}, \omega)$ is the photon annihilation operator of a photon with transverse momentum $\vec{q} = (k_x, k_y)$, 
frequency $\omega$, and polarization $\sigma$.  
The operators satisfy
\begin{align}
\left[ \hat{a}_\sigma (\vec{q}, \omega), \hat{a}^\dagger_{\sigma'} (\vec{q}', \omega') \right] &= \delta_{\sigma, \sigma'} \delta(\vec{q}-\vec{q}') \delta (\omega - \omega').
\end{align}
In addition, $\vec{e}_{\vec{q}, \sigma}$ is the polarization vector, 
and the transverse position in the image plane is $\vec{\rho} = (x,y)$.  
We employ the small bandwidth approximation, where we evaluate the frequency-dependent prefactor at the center frequency $\omega_0$ of the traveling light pulse. We then have $k_z (\omega_0) = n_0 \omega_0 / c$ with $n_0$ the refractive index of the medium, and we further approximate $\vec{e}_{\vec{q}, \sigma} \simeq \vec{e}_{\sigma}$. 
The electric field operator becomes simply 
\begin{align} \label{eq.E-operator}
\vec{E}^{(+)} (\vec{\rho}, t) &=i E_0 \sum_\sigma \vec{e}_{\sigma} \int_0^\infty \!\!d\omega \; \int \!\! d^2 q \; \hat{a}_\sigma (\vec{q}, \omega) e^{i (\vec{q}\cdot\vec{\rho}  - \omega t)},
\end{align}
where
$E_0 = (2\pi)^{-3/2} ( (\hbar \omega_0) / (2 \epsilon_0 n_0 c) )^{1/2}$.

\subsection{Light-matter interaction}

The dipolar light-matter interaction Hamiltonian (in the interaction picture with respect to molecular and field Hamiltonian) is given by
\begin{align} \label{eq.H_l-m}
H_{l-m} (t) &= \int_{\mathds{R}^3} \!\! d^3 r \; \vec{\hat{d}} (t) \cdot \left( \vec{\hat{E}}^{(+)} (\vec{r}, t) + \vec{\hat{E}}^{(-)} (\vec{r}, t) \right),
\end{align}
where $\vec{\hat{d}} (\vec{r}, t) = \vec{\hat{d}}^{(+)} (\vec{r}, t + \vec{\hat{d}}^{(-)} (\vec{r}, t$ are the molecular dipole operator and its positive (negative) frequency components, respectively. 
We consider the interaction of (quantum) light with a few-level sample system, as given by Fig.~\ref{fig.setup}(b). 
It is described by the Hamiltonian
\begin{align}
\hat{H}_{mol} &= \hbar \omega_g \vert g\rangle\langle g \vert +  \hbar \omega_e \vert e\rangle\langle e \vert + \hbar \omega_f \vert f\rangle\langle f \vert, \label{eq.H_mol}
\end{align}
where $\vert g \rangle$ is the ground state. In the following, we set $\omega_g = 0$ without loss of generality. 
The Hamiltonian then consists of an intermediate state $\vert e \rangle$ and final state $\vert f \rangle$.
We can now expand the dipole operator in these states, such that it reads in the Schr\"odinger picture
\begin{align}
\hat{\vec{d}} (\vec{r}) &= \delta (\vec{r} - \vec{r}_{mol}) \left( \vec{d}_{eg} \vert e\rangle \langle g \vert + \vec{d}_{ef} \vert f\rangle \langle e \vert + H. c. \right).
\end{align}
Here $\vec{r}_{mol} = (\vec{\rho}_{mol}, z_{mol})$ denotes the position of the molecule. We assumed it to be point-like, which is an excellent approximation for molecular light-matter interactions in the optical regime.
We further divide it into the positive and negative frequency components,
\begin{align}
\hat{\vec{d}} (\vec{r}) &= \delta (\vec{r} - \vec{r}_{mol}) \left( \hat{\vec{d}}^{(+)} + \hat{\vec{d}}^{(-)} \right),
\end{align}
where
\begin{align}
\hat{\vec{d}}^{(-)} &=  \left( \vec{d}_{eg} \vert e\rangle \langle g \vert + \vec{d}_{ef} \vert f\rangle \langle e \vert \right)
\end{align}
accounts for the excitation of the molecule (and $\vec{d}^{(+)}$ for the de-excitation).

\subsection{ The PDC Model }

In the sketch in Fig.~\ref{fig.setup}(a), a laser pulse with frequency $\omega_p$, bandwidth $\Omega_p$ and transverse beam area $\sim (2\pi /Q_p)^2$ induces spontaneous downconversion in a nonlinear crystal, generating two quantum-correlated beams, traditionally called signal and idler.
Since the purpose of this paper is not to discuss the generation of PDC light in a multimode regime in great detail (many excellent reviews can be found on this topic, e.g. Ref.~\onlinecite{Quesada2022}), here we only introduce the bigaussian model, which allows us to carry out many of the following calculations analytically. 
Crucially, this model includes both spatial and spectral degrees of freedom, and is therefore capable of describing spatiotemporal correlations~\cite{Brambilla2004, Law2004, Gatti2012, Gatti2023} - sometimes dubbed ''X-entanglement"~\cite{Gatti2009} when the two degrees of freedom are not separable. 

\subsubsection{Effective action}
The effective action, which governs the downconversion process in the setup of Fig.~\ref{fig.setup}(a), can be written as~\cite{LaVolpe2021}
\begin{align} \label{eq.H_eff}
\Sigma_{eff}
= \hbar \Gamma \int d\omega_s \int d\omega_i& \; \int d^2 q_s \int d^2q_i \; F_{PDC} ( \vec{q}_s, \omega_s; \vec{q}_i, \omega_i) \notag \\
&\times \hat{a}^\dagger_s (\vec{q}_s, \omega_s) \hat{a}^\dagger_i (\vec{q}_i, \omega_i) + H.c,
\end{align}
Here, the polarization index $\sigma$ of Eq.~(\ref{eq.E-operator}) is replaced by the signal and idler fields, labelled "$s$" and "$i$", respectively. These two fields are assumed to have a fixed polarization, which may be different (in type-II PDC) or identical (in type-I PDC). In the latter case, we assume that the two fields may still be distinguished, e.g. by their propagation direction. 
$\Gamma$ is a dimensionless number, which quantifies the strength of the PDC process. 
We have further introduced the joint spatiotemporal amplitude (JSA) $F_{PDC}$, which we choose to be normalized, i.e.
\begin{align} \label{eq.normalization}
\int d\omega_s \int d\omega_i \; \int d^2 q_s \int d^2q_i \; \left| F_{PDC} ( \vec{q}_s, \omega_s; \vec{q}_i, \omega_i) \right|^2 &= 1.
\end{align}
In general, the JSA has a complicated structure which cannot be factorized into spatial and spectral components~\cite{Brambilla2004}, and which gives rise to high-dimensional $X$-entanglement~\cite{Gatti2009}. 
In this manuscript, we will use a particularly convenient model, where the spatial and the frequency phase matching functions are approximated by Gaussians. This is, e.g., appropriate for non-collinear PDC~\cite{LaVolpe2021}. The JSA then reads
\begin{align} \label{eq.F_PDC}
F_{PDC} (\vec{q}_s, \omega_s; \vec{q}_i, \omega_i) &= F_{mom} (\vec{q}_s, \vec{q}_i) F_{spec} (\omega_s, \omega_i)
\end{align}
with the momentum distribution
\begin{align}
F_{mom} (\vec{q}_s, \vec{q}_i) &= \frac{1}{\pi Q_p Q_m} \exp \left[ - \frac{ (\vec{Q}_s + \vec{Q}_i)^2 }{4 Q_p^2} - \frac{ ( \vec{Q}_s - \vec{Q}_i )^2 }{ 4 Q_m^2 } \right] \label{eq.F_mom}
\end{align}
and the joint spectral amplitude
\begin{align}
F_{spec} (\omega_s, \omega_i) &= \frac{1}{ \sqrt{\pi \Omega_m \Omega_p} } \exp \left[ - \frac{( \Omega_s + \Omega_i )^2}{4 \Omega_p^2} - \frac{ (\Omega_s - \Omega_i)^2 }{ 4 \Omega_m^2 } \right]. \label{eq.F_spec}
\end{align}
Here we have introduced variables that are shifted with respect to their mean, i.e. $\vec{Q}_j = \vec{q}_j - \vec{q}_j^{(0)}$ and $\Omega_j = \omega_j - \omega_j^{(0)}$. 
Due to energy and momentum conservation, we have $\omega_s^{(0)} + \omega_i^{(0)} = \omega_p$ and $\vec{q}_s^{(0)}+\vec{q}_i^{(0)} = 0$. In the following, we will work with degenerate downconversion with $\omega_s^{(0)} = \omega_i^{(0)} = \omega_p / 2$. This reduces some clutter in our notation, but has no implications for the TPA cross section with off-resonant intermediates states, which we will evaluate.
In Eq.~(\ref{eq.F_mom}), the first term in the exponential governs the distribution of the sum of the two momenta, described by the width $Q_p$. 
This parameter is inversely proportional to the transverse beam width $w_0$ of the laser pulse triggering the downconversion process. 
In the following, we will characterize the transverse width of the entangled beam by $A_p = \lambda_p^2 = (2\pi)^2 / Q_p^2$. 
The second term in Eq.~(\ref{eq.F_mom}) describing the distribution of the frequency differences is characterized by the quantity $Q_m$, which is determined by the phase matching in the nonlinear crystal~\cite{Brambilla2004, Beltran2017} and the emission angles of the PDC light which can be collected by the optical system~\cite{LaVolpe2021}. 
Eq.~(\ref{eq.F_spec}) is governed by similar physics: the distribution of the sum of the two photon frequencies is governed by $\Omega_p$, which is the bandwidth of the pump pulse generating the PDC light. The distribution of the frequency differences $\Omega_m$ is again related to the phase matching conditions in the nonlinear crystal. 

The bigaussian form of Eqs.~(\ref{eq.F_mom}) and (\ref{eq.F_spec}) is very convenient, since it allows us to diagonalize the effective action~(\ref{eq.H_eff}) analytically. This can be done with the help of Mehler's formula for Hermite polynomials, which yields~\cite{LaVolpe2021, Horoshko2019} 
\begin{align} \label{eq.Mehler}
\frac{1}{\sqrt{\pi}} \exp \left[ - \frac{1}{4} \frac{1+\zeta}{1-\zeta} (x+y)^2 - \frac{1}{4} \frac{1-\zeta}{1+\zeta} (x-y)^2 \right] \notag \\
= \sqrt{ 1-\zeta^2 } \sum_{n = 0}^\infty (-1)^n \zeta^n h_n (x) h_n (y),
\end{align}
where $0 <\zeta < 1$ and $h_n (k x) = k^{1/2} (2^n n! \sqrt{\pi})^{-1/2} H_n (k x) e^{- (kx)^2 /2}$\footnote{Note that here the Hermite functions are normalized with respect to integrals over space or frequency, i.e. $\int d\omega \; h_n^2 (\omega) = 1$.}. 
Note that the factor $(-1)^n$ is included for the case of Eq.~(\ref{eq.F_mom}) and (\ref{eq.F_spec}), where the first terms in the exponential have a smaller variance than the second terms, i.e. where $Q_p > Q_m$ and $\Omega_p > \Omega_m$. 
In the case where this condition is violated (e.g. when we evaluate the Fourier transform of these functions), we use Eq.~(\ref{eq.Mehler}) without the factor $(-1)^n$.

We find for Eqs.~(\ref{eq.F_mom}) and (\ref{eq.F_spec})
\begin{align} \label{eq.zeta_t}
\zeta_t &= \frac{\Omega_m - \Omega_p}{ \Omega_m + \Omega_p }
\end{align}
and
$\tau = 1 / \sqrt{ \Omega_m \Omega_p }$.
Likewise, for the momentum correlation function we obtain, with $x_j = Q_{sj} q$, $q = 1 / \sqrt{Q_m Q_p}$, and $\zeta_q = (Q_m - Q_p) / (Q_m + Q_p)$. 
Altogether, we can write the PDC action~(\ref{eq.H_eff}) as 
\begin{align}
\Sigma_{eff} &= \hbar \Gamma (1 - \zeta^2_q) \sqrt{1 - \zeta_t^2} \sum_{\vec{n}} \zeta_t^{n_t} \zeta_q^{n_x + n_y} \hat{A}^\dagger_{n_t n_x n_y} \hat{B}^\dagger_{n_t n_x n_y} + h.c.
\end{align}
where we defined the Schmidt mode operators
\begin{align} \label{eq.inputoutput1}
\hat{A}^\dagger_{\vec{n}} &= \int d\omega_s \int d^2q_s \; h_{\vec{n}} (\vec{q}_s, \omega_s) \hat{a}_s^\dagger (\vec{q}_s, \omega_s), 
\end{align}
and
\begin{align} \label{eq.inputoutput2}
\hat{B}^\dagger_{\vec{n}} &= (-1)^{|\vec{n}|} \int d\omega_i \int d^2q_i \; h_{\vec{n}} (\vec{q}_i, \omega_i) \hat{a}_i^\dagger (\vec{q}_i, \omega_i). 
\end{align}
Here we use the short-hand notation $\vec{n} = ({n_t, n_x, n_y})$, and 
\begin{align} \label{eq.Hermite}
h_{\vec{n}} (\vec{q}, \omega) = h_{n_t} \left( \frac{ \omega -  \omega_p/2 }{ \sqrt{ \Omega_m \Omega_p } } \right) h_{n_x} \left( \frac{ q_{x} - q_{x}^{(0)} }{ \sqrt{ Q_m Q_p } } \right) h_{n_y} \left( \frac{  q_{y} - q_{y}^{(0)} }{ \sqrt{ Q_m Q_p } } \right).
\end{align}
The broadband operators inherit their orthonormality from the Hermite functions, i.e. $[\hat{A}_{\vec{n}}, \hat{A}^\dagger_{\vec{n}'}] = \delta_{\vec{n}, \vec{n}'}$,  $[\hat{A}_{\vec{n}}, \hat{A}_{\vec{n}'}] = 0$ (likewise for idler mode $\hat{B}$), and  $[\hat{A}_{\vec{n}}, \hat{B}_{\vec{n}'}] = 0$.

\subsubsection{ Output state }

The quantum state of light generated by the PDC in a nonlinear crystal is given by
\begin{align} \label{eq.psi_BSV}
\vert \psi_{BSV} \rangle &= \exp \left[ - \frac{i}{\hbar} \Sigma_{eff} \right] \vert 0 \rangle.
\end{align}
Correlation functions of this state are most conveniently calculated in the Heisenberg picture, where the input-output relations are readily derived as
\begin{align}
\hat{A}^{(out)}_{\vec{n}} &= \cosh (r_{\vec{n}} \Gamma) \hat{A}_{\vec{n}} + \sinh (r_{\vec{n}} \Gamma) \hat{B}^\dagger_{\vec{n}}, \\
\hat{B}^{(out)}_{\vec{n}}  &= \cosh (r_{\vec{n}} \Gamma) \hat{B}_{\vec{n}} + \sinh (r_{\vec{n}} \Gamma) \hat{A}^\dagger_{\vec{n}},
\end{align}
where $r_{\vec{n}} = (1 - \zeta^2_q) \sqrt{1 - \zeta_t^2} \zeta_t^{n_t } \zeta_q^{n_x + n_y}$. 
Using Eq.~(\ref{eq.E-operator}), the electric field operators in the image plane are then given by 
\begin{align}
\vec{E}_s^{(+)} (\vec{r}, t) &= E_0 \vec{e}_s \sum_{\vec{n}} h_{\vec{n}, s} (\vec{r}, t) \hat{A}_{\vec{n}}^{(out)}, \\
\vec{E}_i^{(+)} (\vec{r}, t) &= E_0 \vec{e}_i \sum_{\vec{n}} h_{\vec{n}, i} (\vec{r}, t) \hat{B}_{\vec{n}}^{(out)}.
\end{align}
Here, 
$E_0$ is defined in Eq.~(\ref{eq.E-operator}) with $\omega_0 = \omega_p/2$, $\vec{e}_j$ the polarization vector, and $h_{\vec{n}, s} (\vec{r}, t)$ ($h_{\vec{n}, i} (\vec{r}, t)$) are the Fourier transforms of the Hermite functions~(\ref{eq.Hermite}), i.e. 
\begin{align} \label{eq.h_n(r)}
h_{\vec{n}, s} (\vec{r}, t) &\equiv \int d\omega_s \int d^2 q_s \; h_{\vec{n}} (\vec{q}_s, \omega_s) e^{i (\vec{q}_s \cdot \vec{\rho} - \omega_s t)},
\end{align}
and 
\begin{align}
h_{\vec{n}, i} (\vec{r}, t) &\equiv (-1)^{|\vec{n}|} \int d\omega_i \int d^2 q_i \; h_{\vec{n}} (\vec{q}_i, \omega_i) e^{i (\vec{q}_i \cdot \vec{\rho} - \omega_i t)}.
\end{align}
Finally, it is possible to calculate the Schmidt number explicitly in this model~\cite{LaVolpe2021}. This quantity measures the effective dimensionality of the state of light.  It is simply given by
\begin{align} \label{eq.Schmidt-number}
K &= K_t K_x K_y,
\end{align}
where we have $K_t = (\Omega_m / \Omega_p + \Omega_p / \Omega_m ) /2$, and $K_x = K_y = (Q_m / Q_p + Q_p / Q_m ) / 2$. 
A useful approximation for the the Schmidt number, which we will use later, is obtained in the limit of strong two-photon entanglement, where e.g. $\Omega_m \gg \Omega_p$ and $Q_m \gg Q_p$, such that the number of spectral modes is given by the ratio of the two numbers, respectively, i.e. $K_t \simeq \Omega_m / (2 \Omega_p)$ and $K_{x/y} \simeq Q_m / (2 Q_p)$. 

\section{Two-photon absorption cross section}
\label{sec.TPA}

\subsection{From the excitation probability to the cross section}

As discussed in detail below, we will use time-dependent perturbation theory to calculate the two-photon excitation probability $P_f (\vec{r}_{mol})$ of a molecule at a position $\vec{r}_{mol}$ interacting with the pulsed quantum light field, which we introduced in the previous section. 
The full TPA signal (which could be measured e.g. by fluorescence) is then given by the sum over all the molecules' contributions, which are contained in a sample. Since this will usually be a macroscopic number, we replace the summation over these contributions by an integral over the sample volume and the average molecular density $m_0$. 
Up to an unimportant constant which depends on the e.g. the detection efficiency of the fluorescence photons, the measured signal is then given by~\footnote{
Note that the factor $1 /  (2\pi)^2$ is introduced for consistency, such that the photon number density $\hat{a}^\dagger (\vec{\rho}, \omega) \hat{a} (\vec{\rho}, \omega)$ integrated over space and frequency yields the mean photon number~(\ref{eq.<N>}). 
}
\begin{align}\label{eq.S_TPA}
S_{TPA} &= \frac{m_0}{(2\pi)^2} \int_{V_{sample}} \!\!\!\! d^3 r_{mol} \; P_{f} (\vec{r}_{mol}).
\end{align}
For the sample volume $V_{sample}$, we consider a thin slab of length $\Delta z$ in the propagation direction of the light, such that we replace $\int dz \simeq \Delta z$.
The transverse extent $\Delta^2 \vec{\rho}$ in contrast is assumed much larger than the transverse width of the light field, such that we can extend the integration boundaries to infinity in the transverse directions. 

The measured TPA rate is then given by the signal and the repetition rate $f_{rep}$ of the laser which generates the entangled pulses, 
\begin{align} \label{eq.R_TPA-def}
R^{TPA} &= S_{TPA} f_{rep}.
\end{align}
To bring this rate into the form of a two-photon absorption rate~(\ref{eq.ETPA-rate_Fei}), we write it as
\begin{align} \label{eq.R^TPA}
R^{TPA} &= N_{mol} \sigma_e \phi + N_{mol} \delta_r \phi^2 
\end{align}
where the total number of molecules in the sample, which are irradiated by the light, is given by $N_{mol} =  m_0 \Delta z A_p$. 
To establish the scaling with the photon flux density, we write the photon flux density as
\begin{align} \label{eq.phi}
\phi &= \langle \hat{N} \rangle \times \frac{f_{rep}}{A_p},
\end{align}
where 
\begin{align} \label{eq.<N>}
\langle \hat{N} \rangle &= \sum_{j = s,i} \int d\omega \int d^2q \; \langle \psi_{BSV} \vert \hat{a}^\dagger_{j} (\vec{q}, \omega) \hat{a}_{j} (\vec{q}, \omega)\vert \psi_{BSV} \rangle \notag \\
&= 2 \sum_{\vec{n}} \sinh^2 (r_{\vec{n}} \Gamma)
\end{align}
is the total photon number during one pulse, 
$f_{rep}$ accounts for the pulse duration as above,

\begin{figure*}
\includegraphics[width=0.8\textwidth]{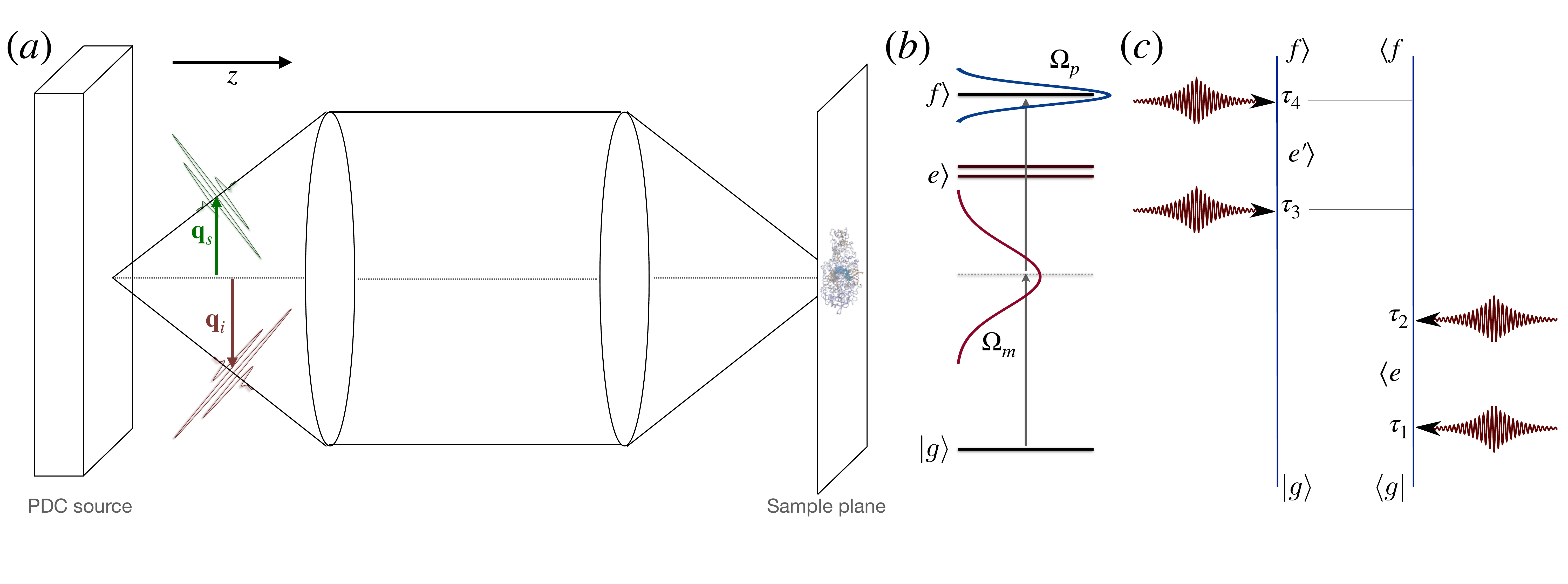}
\centering
\caption{
(a) Schematic setup considered in this manuscript: A thin slab containing the sample molecules is placed in the object plane of an imaging system, such that near-field correlations of the light fields created by parametric downconversion can be harnessed~\cite{Brambilla2004}. (b) Molecular level scheme: the broadband photons generated in the PDC process trigger TPA events in the molecule. The intermediate states $\vert e \rangle$, which are dipole-coupled to the ground state, are far detuned from the centre frequency $\omega_p / 2$ of each photon. The final state $\vert f \rangle$ is near-resonant with the sum of the two photon frequencies $\omega_p$. 
(c) Double-sided Feynman diagram representing Eq.~(\ref{eq.P_f}). For diagram rules, see e.g. Ref.~\onlinecite{Dorfman16}.
}
\label{fig.setup}
\end{figure*}

\subsection{Experimental setup and two-photon absorption probability}

We next calculate the probability of entangled TPA by a point-like molecule at position $\vec{r}_{mol} = (\vec{\rho}_{mol}, 0)$ in the image plane of the optical setup (see Fig.~\ref{fig.setup}(a)). 
The two-photon absorption cross section can be calculated from the probability to excite the final molecular state $f$ via the absorption of two photons. Thus, we need to calculate
\begin{align} \label{eq.P_f-exact}
P_f (\vec{r}_{mol}, t) &= \tr \left\{ \vert f(t) \rangle \langle f(t) \vert \mathcal{T} \exp \left[ - \frac{i}{\hbar} \int^t_{t_0} \!\! d\tau H_{l-m, -} (\tau) \right] \varrho_0 \right\}, 
\end{align}
where the initial state of the light-matter system at time $t_0$ is given by the state of light emitted by the PDC source, Eq.~(\ref{eq.psi_BSV}), and the molecular electronic ground state,
\begin{align}
\varrho_0 &= \varrho_{sys} \otimes \varrho_f, 
\end{align}
with $\varrho_{sys} = \vert g(t_0) \rangle \langle g(t_0) \vert$ and $\varrho_f = \vert \psi_{BSV} (t_0) \rangle \langle \psi_{BSV} (t_0) \vert$. 
The time evolution is given by the Dyson series, and defined the time-ordering operator $\mathcal{T}$. We also defined the light-matter interaction superoperator $H_{l-m, -}$, which acts on the density matrix as $H_{l-m, -} \varrho = H_{l-m} \varrho - \varrho H_{l-m}$. The interaction Hamiltonian in defined in Eq.~(\ref{eq.H_l-m}), and the electric field operator is given by the sum of signal and idler fields, $\vec{E} = \vec{E}_s + \vec{E}_i$.
The final state population can be calculated from Eq.~(\ref{eq.P_f-exact}) by expanding the exponential of the Dyson series, and collecting the leading-order contributions. 

Using the rotating wave approximation, $H_{l-m} \simeq \vec{d}^{(+)} \cdot \vec{E}^{(-)} + \vec{d}^{(-)} \cdot \vec{E}^{(+)}$, and the limiting our analysis to off-resonant intermediate states, we find the probability to excite the f-state population is given by the Feynman diagram in Fig.~\ref{fig.setup}(c) (for details, see e.g. Ref.~\onlinecite{Raymer2021b}), which evaluates to 
\begin{widetext}
\begin{align}
P_f (\vec{r}_{mol}, t) &= \int^t_{t_0} \!\!d\tau_4 \int^{\tau_4}_{t_0}\!\!d\tau_3 \int^{\tau_3}_{t_0} \!\!d\tau_2 \int^{\tau_2}_{t_0} \!\!d\tau_1 \!\!\!\!\!\! \sum_{j_1, j_2, j_3, j_4 = s, i} \!\!\!\!\!\! 2 \Re \bigg( C^{mat}_{j_1, j_2, j_3, j_4} (\tau_1, \tau_2, \tau_3. \tau_4) C^{field}_{j_1, j_2, j_3, j_4} (\vec{r}_{mol}, \tau_1, \vec{r}_{mol}, \tau_2; \vec{r}_{mol}, \tau_3, \vec{r}_{mol}, \tau_4) \bigg), \label{eq.P_f}
\end{align}
\end{widetext}
where
\begin{align}
&C^{mat}_{j_1, j_2, j_3, j_4} (\tau_1, \tau_2, \tau_3. \tau_4) \notag \\
= &\frac{1}{\hbar^4} \tr \left\{ \mu^{(+)}_{j_1} (\tau_1)\mu^{(+)}_{j_2} (\tau_2) \mu^{(-)}_{j_4} (\tau_4) \mu^{(-)}_{j_3} (\tau_3) \varrho_{sys}\right\}
\end{align}
is the matter correlation response function.
We further defined the projections of the molecular dipole operators on the polarization of the incoming light field, 
\begin{align}
\mu_j (t) &= \vec{e}_j \cdot \vec{d} (t). 
\end{align}
Likewise, the correlation function of the field operators is given by
\begin{align}
&C^{field}_{j_1, j_2, j_3, j_4} (\vec{r}_{mol}, \tau_1; \vec{r}_{mol},\tau_2; \vec{r}_{mol}, \tau_3; \vec{r}_{mol}, \tau_4) \notag \\
= &\tr \left\{ E^{(-)}_{j_1} (\vec{r}_{mol}, \tau_1) E^{(-)}_{j_2} (\vec{r}_{mol}, \tau_2) E^{(+)}_{j_4} (\vec{r}_{mol}, \tau_4) E^{(+)}_{j_3} (\vec{r}_{mol}, \tau_3) \varrho_f \right\}.
\end{align}
The impact of the polarization degrees of freedom can be calculated with a rotational average. For ETPA, this was done in Ref.~\cite{Frontiers2022}, and we found that it can account only for minor changes of the ETPA cross section. Thus, in the following, we will neglect this procedure, and replace the dipole operators by scalar quantities, i.e. $\mu_j \rightarrow \mu$. 

\subsubsection{The molecular correlation function}

We will evaluate the molecular correlation function first. 
Using the Feynman diagram in Fig.~\ref{fig.setup}(c), we evaluate the correlation function straightforwardly to find
\begin{align}
C^{mat} (\tau_1, \tau_2, \tau_3. \tau_4) &= \sum_{e, e'} \mu^\ast_{ge} \mu^\ast_{ef} \mu_{ge'} \mu_{e'f} e^{i ( \omega_{eg} + i \gamma_{eg}) (\tau_2 - \tau_1)} \notag \\
&\times e^{i (\omega_{fg} + i \gamma_{fg}) (\tau_3 -\tau_2)} e^{i (\omega_{fe'} + i \gamma_{e'f}) (\tau_4 - \tau_3)}.
\end{align}
This expression includes the summation over all the virtual states $e, e'$, through which the ETPA process can take place. 
We have further added phenomenological dephasing rates $\gamma_{ge}$ and $\gamma_{ef}$, which broaden the linewidths of the resonances. The correlation function is most conveniently evaluated by changing to the time delay integration variables $t_1 = \tau_2 - \tau_1$, $t_2 = \tau_3 - \tau_2$, and $t_3 = \tau_4 - \tau_3$, and sending $t_0 \rightarrow - \infty$. 
We are further interested in the limit $t\rightarrow \infty$, i.e. in the final population after the entangled pulse has interacted with the molecule. 
In this limit, the $\tau_{4}$-integration simply evaluates to $2\pi \delta (\omega_1 + \omega_2 - \omega_3 - \omega_4)$. 
With the definition of the electric field operator~(\ref{eq.E-operator}), we arrive at
\begin{widetext}
\begin{align}
P_f (\vec{r}_{mol}, t\rightarrow\infty) &=  2 (2\pi)^5 \Re \frac{E_0^4}{\hbar^4} \sum_{e, e'} \mu^\ast_{ge} \mu^\ast_{ef} \mu_{ge'} \mu_{e'f} \int_0^{\infty} \!\!\! d\omega_1 \int_0^{\infty} \!\!\! d\omega_2 \int_0^{\infty} \!\!\! d\omega_3 \int_0^{\infty} \!\!\! dt_3 \int_0^{\infty} \!\!\! dt_2 \int_0^{\infty} \!\!\! dt_1 \notag \\
&\times e^{i (\omega_1 -\omega_{eg} + i \gamma_{eg}) t_1 + i (\omega_1 + \omega_2 - \omega_{fg} + i \gamma_{fg}) t_2 + i (\omega_1 + \omega_2 - \omega_3 - \omega_{fe'} + i \gamma_{fe'}) t_3} \notag \\
&\times \sum_{j_1, j_2, j_3, j_4} \tr \left\{ \hat{a}^\dagger_{j_3} (\vec{r}_{mol}, \omega_3) \hat{a}^\dagger_{j_4} (\vec{r}_{mol}, \omega_1+\omega_2 - \omega_3) \hat{a}_{j_2} (\omega_2) \hat{a}_{\vec{r}_{mol}, j_1} (\vec{r}_{mol}, \omega_1) \varrho_f \right\}. 
\end{align}
\end{widetext}
Here, we use the mixed representation of the field operators,
\begin{align} \label{eq.a(r)}
\hat{a} (\vec{r}, \omega) &= \frac{1}{2\pi} \int_{\mathds{R}^2} d^2 q \; \hat{a} (\vec{q}, \omega) \; e^{i (\vec{q} \cdot\vec{\rho} + k_z (\omega) z)}. 
\end{align}
Evaluating the time integrals for the case of off-resonant intermediate states, we replace
\begin{align}
\int_0^{\infty} \!\! dt_1 \; e^{i (\omega_1 -\omega_{eg} + i \gamma_{eg} ) t_1} &= \frac{i}{\omega_1 - \omega_{eg} + i \gamma_{eg}} \simeq \frac{i}{\omega_p/2 - \omega_{eg} },
\end{align}
and likewise for the $t_3$-integration, where we find
\begin{align}
\int_0^{\infty} \!\! dt_3 \; e^{ i (\omega_1 + \omega_2 - \omega_3 - \omega_{fe'} + i \gamma_{fe'}) t_3} \simeq  \frac{i}{\omega_p/2 - \omega_{fe'} }.
\end{align}
This approximation assumes that the photon bandwidth $\sim \Omega_m$ is much smaller than the detuning to the electronic resonances, as indicated in Fig.~\ref{fig.setup}(b). 
We then obtain
\begin{widetext}
\begin{align} \label{eq.P_f-final}
P_f (\vec{r}_{mol}, t\rightarrow\infty) &= \frac{1}{4} \left( \frac{\omega_p / 2}{\hbar \epsilon_0 n_0 c} \right)^2 \sum_{e, e'} \frac{\mu^\ast_{ge} \mu^\ast_{ef} \mu_{ge'} \mu_{e'f}}{ (\omega_{eg} - \omega_p/2) (\omega_p/2 - \omega_{fe'}) } \int_0^{\infty} \!\!\! d\omega_1 \int_0^{\infty} \!\!\! d\omega_2 \int_0^{\infty} \!\!\! d\omega_3 \frac{1}{\pi} \frac{\gamma_{fg}}{\gamma_{fg}^2 + (\omega_{fg} - \omega_1 - \omega_2)^2} \notag \\
&\times \sum_{j_1, j_2, j_3, j_4} \tr \left\{ \hat{a}^\dagger_{j_3} (\vec{r}_{mol}, \omega_3) \hat{a}^\dagger_{j_4} (\vec{r}_{mol}, \omega_1+\omega_2 - \omega_3) \hat{a}_{j_2} (\omega_2) \hat{a}_{\vec{r}_{mol}, j_1} (\vec{r}_{mol}, \omega_1) \varrho_f \right\}.
\end{align}
Here we have taken the real part explicitly, using that the molecular response encapsulated in the term $\sum_{e, e'} \ldots$ is real. 

\subsubsection{The field correlation function}

We next turn to the remaining field correlation function, where we can use the input output relations~(\ref{eq.inputoutput1}) and (\ref{eq.inputoutput2}) to obtain
\begin{align} \label{eq.field-correlation}
&\sum_{j_1, j_2, j_3, j_4} \tr \left\{ \hat{a}^\dagger_{j_3} (\vec{r}_{mol}, \omega_3) \hat{a}^\dagger_{j_4} (\vec{r}_{mol}, \omega_4) \hat{a}_{j_2} (\omega_2) \hat{a}_{\vec{r}_{mol}, j_1} (\vec{r}_{mol}, \omega_1) \varrho_f \right\} \notag \\
= &f^\ast (\vec{r}_{mol}, \omega_1, \omega_2) f(\vec{r}_{mol}, \omega_3, \omega_4) + g (\vec{r}_{mol}, \omega_1, \omega_3) g (\vec{r}_{mol}, \omega_2, \omega_4) + g (\vec{r}_{mol}, \omega_1, \omega_4) g (\vec{r}_{mol}, \omega_2, \omega_3),
\end{align}
where $\omega_4 = \omega_1+\omega_2 - \omega_3$ and we defined
\begin{align}\label{eq.corr-corr-fct}
f (\vec{r}, \omega_1, \omega_2) &= e^{ - i \omega_p t} \sum_{\vec{n}} \sinh (r_{\vec{n}} \Gamma) \cosh (r_{\vec{n}} \Gamma) \left( h_{\vec{n}, s} (\vec{r}, \omega_1) h_{\vec{n}, i} (\vec{r}, \omega_2) + h_{\vec{n}, i} (\vec{r}, \omega_1) h_{\vec{n}, s} (\vec{r}, \omega_2) \right)
\end{align}
and
\begin{align}\label{eq.unc-corr-fct}
g (\vec{r}, \omega_1, \omega_2) &= \sum_{\vec{n}} \sinh^2 (r_{\vec{n}} \Gamma) \left( h_{\vec{n}, s} (\vec{r}, \omega_1) h_{\vec{n}, s} (\vec{r}, \omega_2) + h_{\vec{n}, i} (\vec{r}, \omega_1) h_{\vec{n}, i} (\vec{r}, \omega_2) \right).
\end{align}
The Hermite functions of the signal mode (and similarly the idler modes) are given by
\begin{align} \label{eq.h_n(r,omega)}
h_{\vec{n}, s} (\vec{r}, \omega) &\equiv \frac{1}{2\pi} \int_{\mathds{R}^2} d^2 q_s \; h_{\vec{n}} (\vec{q}_s, \omega_s) e^{i \vec{q}_s\cdot \vec{\rho} } \notag \\
&=  i^{n_x+n_y} e^{ i \vec{q}_s^{(0)} \vec{\rho} } \; h_{n_t} \left( \frac{ \omega - \omega_p /2 }{ \sqrt{\Omega_m \Omega_p} } \right) h_{n_x} \left(\sqrt{Q_mQ_p} x\right) h_{n_y} \left(\sqrt{Q_mQ_p} y \right).
\end{align}
\end{widetext}

The function $f$ describes the absorption of a correlated photon pair, as it contains only pairs of signal and idler functions, $h_{\vec{n}, s}$ and $h_{\vec{n}, i}$, respectively, whereas the function $g$ corresponds to the sum of the autocorrelation contributions of the two beams. 
We note that spatial and spectral correlations, which are encoded in the Schmidt numbers $r_{\vec{n}}$, are connected non-trivially in Eq.~(\ref{eq.field-correlation}). The degrees of freedom can be separated only in the weak downconversion limit $r_{\vec{n}}\Gamma \ll 1$, when the light field is composed predominantly of temporally separate entangled photon pairs. 
In the following, we will therefore consider three limiting case: first, we consider the two-photon limit when the correlations are in fact separable. Second, when the light field is confined to a single spatial mode and only spectral correlations matter, and, third, when there is only a single spectral mode, and spatial correlations affect the signal. 

\subsubsection{The full two-photon excitation probability}

Inserting the field correlation function~(\ref{eq.field-correlation}) into Eq.~(\ref{eq.P_f-final}), we find two contributions to the excitation probability, 
\begin{align}\label{eq.P_f-2terms}
P_f (\vec{r}_{mol}, t\rightarrow\infty) &= P_f^{corr} + P_f^{unc},
\end{align}
where $ P_f^{corr}$ stems from the absorption of correlated pairs as described by Eq.~(\ref{eq.corr-corr-fct}) and $ P_f^{unc}$ from absorption according to Eq.~(\ref{eq.unc-corr-fct}).

\section{The two-photon limit}
\label{sec.two-photons}

We first analyze the two-photon limit of the output state~(\ref{eq.psi_BSV}). When $r_{\vec{n}} \Gamma \ll 1$, we can approximate
$\vert \psi_{BSV} \rangle \simeq \vert 0 \rangle + \Gamma \ldots \hat{a}_s^\dagger \hat{a}_i^\dagger \vert 0\rangle$. The field correlation function~(\ref{eq.field-correlation}) simplifies, using Eq.~(\ref{eq.Mehler}), to
\begin{align} \label{eq.f_low-gain}
f_{low-gain} (\vec{r}_{mol}, \omega_1, \omega_2) &\simeq 2 \times \Gamma F_{mom} (\vec{\rho}_{mol}, \rho_{mol}) F_{spec} (\omega_1, \omega_2),
\end{align}
where the Fourier transform of the momentum correlation function evaluated at the position of the molecule $\vec{\rho}_{mol}$ reads
\begin{align} \label{eq.F_mom(r)}
&F_{mom} (\vec{\rho}_{mol}, \rho_{mol}) \notag \\
&= \frac{1}{(2\pi)^2} \int_{\mathds{R}^2} d^2 q_s \int_{\mathds{R}^2} d^2 q_i \; F_{mom} (\vec{Q}_s, \vec{Q}_i) e^{i (\vec{q}_s+\vec{q}_i) \vec{\rho}_{mol}} \notag \\
&= \frac{1}{\pi} Q_m Q_p \exp \left[ - Q_p^2 \vec{\rho}_{mol}^2 + i (\vec{q}_s^{(0)} + \vec{q}_i^{(0)} ) \cdot\vec{\rho}_{mol} \right].
\end{align}
The factor 2 at the beginning of Eq.~(\ref{eq.f_low-gain}) stems from the fact that the correlation functions~(\ref{eq.corr-corr-fct}) are symmetric with respect to the exchange $\omega_s \leftrightarrow \omega_i$ and $\vec{Q}_s \leftrightarrow \vec{Q}_i$ (the signal photon or the idler photon can be absorbed first, respectively). 
These two excitation pathways contribute the same result and yield the overall factor 2. 
Hence, the TPA rate~(\ref{eq.R_TPA-def}) separates into a spatial and frequency integrals, which can be treated separately, 
\begin{align} \label{eq.R^TPA_low-gain}
R^{TPA}_{low-gain} &= r_0 p_{spat} p_{freq}.
\end{align}
The prefactor $r_0$ can be read off from Eqs.~(\ref{eq.P_f-final}) and (\ref{eq.R_TPA-def}),
\begin{align}
r_0 &= m_0 f_{rep} \Gamma^2 \left( \frac{\omega_p / 2}{ \hbar \epsilon_0 n_0 c} \right)^2 \sum_{e, e'} \frac{\mu_{ge}^\ast \mu_{ef}^\ast \mu_{e'f} \mu_{g e'}}{ (\omega_{eg} - \omega_p/2) (\omega_p/2 - \omega_{fe'}) }.
\end{align}
The spatial contribution $p_{spat}$, given by the modulus square of Eq.~(\ref{eq.F_mom(r)}) integrated over the transverse position $\rho$, readily evaluates to
\begin{align}
r_{spat}  &= \frac{\Delta z}{ (2\pi)^2 } \int_{\mathds{R}^2} d^2 \rho_{mol} \left( \frac{Q_m Q_p}{\pi} \right)^2 e^{- 2 Q_p^2 (x^2 + y^2)} \notag  \\
& =  \frac{\Delta z Q_m^2}{2\pi},
\end{align}
and finally the frequency contribution is given by
\begin{widetext}
\begin{align} 
p_{freq} &= \int_0^{\infty} \!\!\! d\omega_1 \int_0^{\infty} \!\!\! d\omega_2 \int_0^{\infty} \!\!\! d\omega_3 \frac{1}{\pi}\frac{ \gamma_{fg} }{ \gamma_{fg}^2 + (\omega_{fg} - \omega_1 - \omega_2)^2 } F_{spec} (\omega_3, \omega_1+\omega_2 - \omega_3) F_{spec} (\omega_1, \omega_2). 
\end{align}
\end{widetext}
This last dimensionless quantity encodes the enhancement of the ETPA cross section due to frequency entanglement. It was analyzed by Raymer et al. in great detail~\cite{Raymer2020, Landes2021, Landes2021b, Raymer2021, Raymer2021b}, and depends crucially on the broadening $\gamma_{fg}$ of the final molecular state. If the broadening is very small, as proposed by Kang et al.~\cite{Kang2020} for ETPA in organic chromophores, one obtains a very large enhancement. If it is as broad as suggested by Raymer and coworkers for molecules in solution, the enhancement due to spectral entanglement is almost completely eroded. 
With the bigaussian model for the JSA~(\ref{eq.F_spec}), we can solve the above integrations analytically, when we extend the integration boundaries to $-\infty$. 
In case of resonant excitation, where $\omega_p = \omega_{fg}$, we obtain
\begin{align}\label{eq.p_freq}
p_{freq} &=  \frac{\Omega_m}{\Omega_p} \; \text{erfcx} \left( \frac{\gamma_{fg}}{\sqrt{2}\Omega_p} \right),
\end{align}
where we find the scaled complementary error function erfcx.
It quantifies how efficiently the light field can activate the full two-photon response of the molecule. 
\begin{figure}
\includegraphics[width=0.45\textwidth]{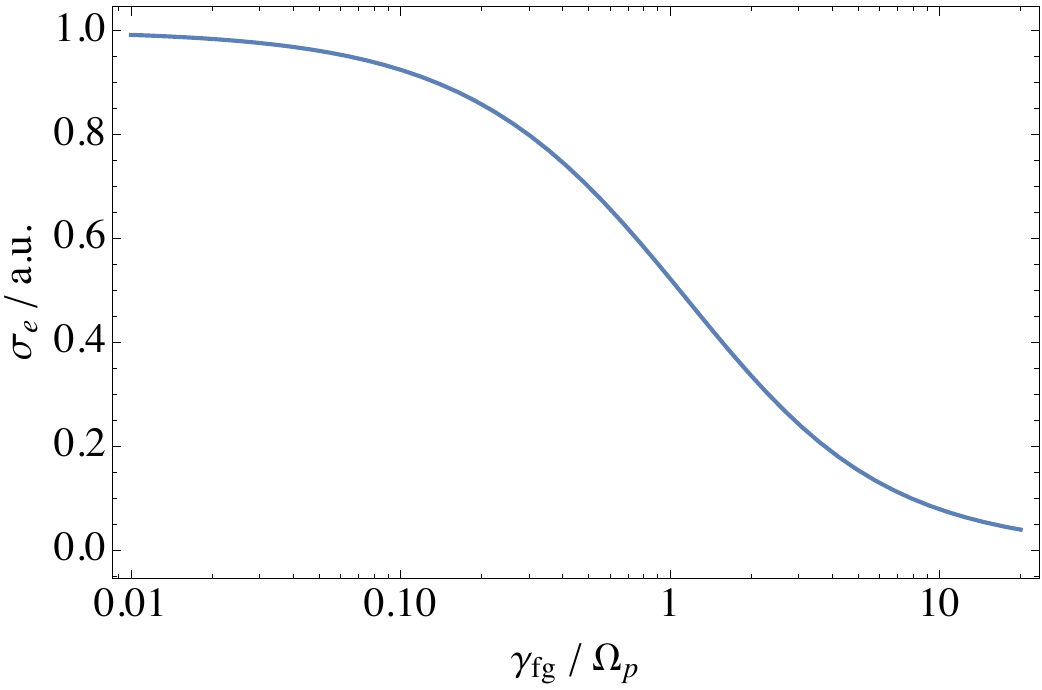}
\centering
\caption{
The $\gamma_{fg}$-dependence of the ETPA cross section, Eq.~(\ref{eq.sigma_e(gamma)}), is shown as a function of $\gamma_{fg}$ in units of the pump bandwidth $\Omega_p$. 
}
\label{fig.sigma_e}
\end{figure}

To obtain the photon flux scaling  from Eq.~(\ref{eq.R^TPA_low-gain}), and hence extract the ETPA cross section, we evaluate the photon flux density in the low-gain regime
$\phi_{low-gain} \simeq \frac{f_{rep}}{A_p} 2 \Gamma^2 \sum_{\vec{n}} r^2_{\vec{n}} =  2 \frac{f_{rep}}{A_p}  \Gamma^2$.
The second equality follows from the geometric series (or, equivalently, from the normalization~(\ref{eq.normalization}) of the JSA). 
With Eq.~(\ref{eq.R^TPA}), we then obtain the ETPA absorption rate
\begin{align} \label{eq.sigma_e}
\sigma_e 
&= \frac{\sigma^{(2)}}{A_e T_e} \text{eff} (\gamma_{fg} / \Omega_p). 
\end{align}
In the first term, we encounter the sought-after result~(\ref{eq.ETPA-cross-section_Fei}) with the classical TPA cross section~\cite{Landes2021}
\begin{align} \label{eq.sigma^2}
\sigma^{(2)} &= \left( \frac{\omega_p/2}{ \hbar \epsilon_0 n c } \right)^2 \frac{1}{2\gamma_{fg} } \sum_{e, e'} \frac{\mu_{ge}^\ast \mu_{ef}^\ast \mu_{e'f} \mu_{g e'}}{ (\omega_{eg} - \omega_p/2) (\omega_p/2 - \omega_{fe'}) },
\end{align}
as well as the entanglement area 
\begin{align} \label{eq.A_e}
A_e &\equiv \frac{(2\pi)^2}{Q_m^2}
\end{align}
and the entanglement time 
\begin{align} \label{eq.T_e}
T_e &\equiv \frac{2\pi}{\Omega_m}. 
\end{align}
In addition, the 'classical' formula~(\ref{eq.ETPA-cross-section_Fei}) is multiplied by the efficiency function
\begin{align}\label{eq.efficiency}
\text{eff} (x) &\equiv x \;  \text{erfcx} \left( \frac{x}{\sqrt{2}} \right). 
\end{align}
At large $x$, when $\gamma_{fg} \gg \Omega_p$, the function saturates to eff$(x) \simeq 0.8$. 
This situation may be encountered in molecular TPA with cw entangled photons, where you can simply use
\begin{align}
\sigma^{mol}_e &\simeq \frac{\sigma^{(2)}}{ A_e T_e},
\end{align}
in accordance with Fei et al.~(\ref{eq.ETPA-cross-section_Fei}). 
At small $x$, in contrast, i.e. when $\gamma_{fg} \ll \Omega_p$, we find eff$(x) \simeq x$. 
In this limit, the pump bandwidth $\Omega_p$ is so large that it limits the efficient excitation of the molecule. 
It is important to keep in mind, however, that the classical TPA cross section is inversely proportional to the resonance broadening, $\sigma^{(2)} \propto 1/\gamma_{fg}$. Therefore, the above discussion must be understood at fixed $\gamma_{fg}$, and for varying pump bandwidth $\Omega_p$. In particular, it does \textit{not} imply that a large broadening that a large broadening, which pushes eff$(x)$ towards saturation, would be beneficial for ETPA. Instead, if we include the full dependence of $\sigma_e$ on the broadening, we find that 
\begin{align} \label{eq.sigma_e(gamma)}
\sigma_e (\gamma_{fg}) \propto \frac{1}{\gamma_{fg}} \text{eff} \left( \frac{\gamma_{fg}}{ \sqrt{2} \Omega_p } \right).
\end{align}
As shown in Fig.~\ref{fig.sigma_e}, this function only decreases with increasing broadening $\gamma_{fg}$. 
When $\gamma_{fg} \ll \Omega_p$, we have eff$(x) \simeq x$, such that the broadening dependence cancels and we obtain an ETPA cross section with $\gamma_{fg}$ in Eq.~(\ref{eq.sigma^2}) replaced by $\Omega_p$. 
When $\gamma_{fg} \gg \Omega_p$, Eq.~(\ref{eq.efficiency}) becomes constant, such that $\sigma_e \propto 1 / \gamma_{fg}$. 

\section{Single spatial mode limit}
\label{sec.single-spatial-mode}
When we have $Q_m = Q_p$, signal and idler beams each propagate in one respective spatial mode, but may still contain strong spectral quantum correlations. 
In this case, the momentum correlation function factorizes naturally into 
\begin{align}
F_{mom}^{sep} (\vec{q}_s, \vec{q}_i) &= \frac{1}{\pi Q_p^2} \exp \left[ - \frac{ (\vec{Q}_s + \vec{Q}_i)^2 }{4 Q_p^2} - \frac{ ( \vec{Q}_s - \vec{Q}_i )^2 }{ 4 Q_p^2 } \right] \\
&=\frac{1}{\pi Q_p^2} \exp \left[ - \frac{\vec{Q}_s^2 + \vec{Q}_i^2}{2 Q_p^2} \right].
\end{align} 
Consequently, we can drop the $n_x / n_y$-summations in the field correlation functions~(\ref{eq.corr-corr-fct}) and (\ref{eq.unc-corr-fct}), and replace the corresponding Hermite functions by 
\begin{align}
\tilde{h} (\vec{r}) &\equiv \frac{Q_p}{\sqrt{\pi}} e^{- i \vec{q}^{(0)} \vec{\rho} - Q_p^2 \vec{\rho}^2 /2 }.
\end{align}
The field correlation functions are thus given by
\begin{widetext}
\begin{align}
f_{1-mode} (\vec{r}, \omega_1, \omega_2) &= 2\sum_{n_t} \sinh (r_{n_t} \Gamma) \cosh (r_{n_t} \Gamma) \tilde{h}^2 (\vec{r}) (-1)^{n_t} h_{n_t} \left( \frac{\omega_p/2 + \omega_1}{ \sqrt{\Omega_m \Omega_p} }\right) h_{n_t} \left( \frac{\omega_p/2 + \omega_2}{ \sqrt{\Omega_m \Omega_p} }\right),
\end{align}
and
\begin{align}
g_{1-mode} (\vec{r}, \omega_1, \omega_2) &= 2\sum_{n_t} \sinh^2 (r_{n_t} \Gamma) | \tilde{h} (\vec{r}) |^2  h_{n_t} \left( \frac{\omega_p/2 + \omega_1}{ \sqrt{\Omega_m \Omega_p} }\right) h_{n_t} \left( \frac{\omega_p/2 + \omega_2}{ \sqrt{\Omega_m \Omega_p} }\right).
\end{align}
The photon flux density becomes
\begin{align} \label{eq.photon-flux_single-spatial-mode}
\phi &= 2\frac{f_{rep}}{A_p}  \sum_{n_t} \sinh^2 (r_{n_t} \Gamma). 
\end{align}

\subsection{Correlated contribution}
We then find from Eq.~(\ref{eq.P_f-final}) the correlated contribution of the final state population
\begin{align} \label{eq.P_f^corr1}
P_f^{corr} (\vec{r}_{mol}) &= \left( \frac{\omega_p / 2}{\hbar \epsilon_0 n_0 c} \right)^2 \sum_{e, e'} \frac{\mu^\ast_{ge} \mu^\ast_{ef} \mu_{ge'} \mu_{e'f}}{ (\omega_{eg} - \omega_0) (\omega_0 - \omega_{fe'}) } \tilde{h}^4 (\vec{r}_{mol}) \sum_{n_t, n_t'} (-1)^{n_t + n'_t} \sinh (r_{n_t} \Gamma) \cosh (r_{n_t} \Gamma) \sinh (r_{n_t'} \Gamma) \cosh (r_{n_t'} \Gamma) \notag \\
&\times \int_0^{\infty} \!\!\! d\omega_{sum} \int_0^{\infty} \!\!\! d\omega_2 \int_0^{\infty} \!\!\! d\omega_3 \frac{1}{\pi} \frac{\gamma_{fg}}{\gamma_{fg}^2 + (\omega_{fg} - \omega_{sum})^2 }  h_{n_t} (\omega_{sum}-\omega_2) h_{n_t} (\omega_2)  h_{n'_t} (\omega_3) h_{n'_t} (\omega_{sum}-\omega_3).
\end{align}
When we extend the lower integration boundaries to $-\infty$, and use the parity of the Hermite functions, $h_n (\omega_{sum} - \omega) = (-1)^n h_n (\omega - \omega_{sum})$, we can use the integral identity~\cite{Cahill1969}
\begin{align}
\int_{-\infty}^\infty \!\!\!\! d\omega \; h_m (\omega) h_n (\omega_{sum} - \omega) 
&= \sqrt{\frac{n!}{m!}} \delta\omega^{m-n}  L^{(m-n)}_n \left(\delta\omega^2 \right) e^{- \delta\omega^2 / 2}, \quad \text{for } m \geq n, \label{eq.Laguerre}
\end{align}
where $L^{(m-n)}_n$ is the n-th generalized Laguerre polynomial and $\delta\omega = (\omega_{sum} - \omega_p) / \sqrt{2 \Omega_m \Omega_p }$, 
and $w = (\omega_{sum} - \omega_p) / \sqrt{2\Omega_m \Omega_p}$. Eq.~(\ref{eq.P_f^corr1}) evaluates to
\begin{align} \label{eq.P_corr-exact}
P_f^{corr} (\vec{r}_{mol}) &=  \left( \frac{\omega_p / 2}{\hbar \epsilon_0 n_0 c} \right)^2 \sum_{e, e'} \frac{\mu^\ast_{ge} \mu^\ast_{ef} \mu_{ge'} \mu_{e'f}}{ (\omega_{eg} - \omega_0) (\omega_0 - \omega_{fe'}) } \tilde{h}^4 (\vec{r}_{mol}) \sum_{n_t, n_t'} \sinh (r_{n_t} \Gamma) \cosh (r_{n_t} \Gamma) \sinh (r_{n_t'} \Gamma) \cosh (r_{n_t'} \Gamma) \notag \\
&\times \int_{w_0}^{\infty} \!\! dw \frac{1}{\pi} \frac{w_{fg}}{w_{fg}^2 + (\frac{\omega_{fg} - \omega_p}{\sqrt{2\Omega_m\Omega_p}} + w)^2} L_{n_t} \left( w^2 \right) L_{n_t'} \left( w^2 \right) \exp \left[ - w^2 \right]
\end{align}
\end{widetext}
where $w_{fg} = \gamma_{fg} / \sqrt{2\Omega_m\Omega_p}$ and $w_0 = - \omega_p / \sqrt{2\Omega_p \Omega_m}$. 
Sending $w_0 \rightarrow -\infty$, the remaining $w$-integration can be carried out analytically or numerically by Mathematica with high efficiency (albeit the integration becomes unstable for extremely highly entangled states). 
Still, it is instructive to consider one limiting case first. 

\paragraph{Narrow resonance}
In the limit of a very narrow resonance, when $w_{fg} \ll 1$, we can replace 
\begin{align} \label{eq.delta-approximation}
\frac{1}{\pi} \frac{w_{fg}}{w_{fg}^2 + (\frac{\omega_{fg} - \omega_p}{\sqrt{2\Omega_m\Omega_p}} + w)^2} &\simeq \delta \left(w + \frac{\omega_{fg} - \omega_p}{\sqrt{2\Omega_m\Omega_p}} \right).
\end{align} 
We then obtain on resonance ($\omega_{fg} = \omega_p$) from Eq.~(\ref{eq.P_corr-exact})
\begin{align} \label{eq.P_f-delta}
P_f^{corr} (\vec{r}_{mol}) &= \left( \frac{\omega_p / 2}{\hbar \epsilon_0 n_0 c} \right)^2 \sum_{e, e'} \frac{\mu^\ast_{ge} \mu^\ast_{ef} \mu_{ge'} \mu_{e'f}}{ (\omega_{eg} - \omega_0) (\omega_0 - \omega_{fe'}) } \tilde{h}^4 (\vec{r}_{mol}) \notag \\
&\times \left[ \sum_{n_t} \sinh (r_{n_t} \Gamma) \cosh (r_{n_t} \Gamma) \right]^2
\end{align}
The expression can be interpreted in two limiting cases: At a low photon flux, when $\Gamma \ll 1$, we can approximate, using the definition of the Schmidt eigenvalues in Eq.~(\ref{eq.Mehler}), 
\begin{align}
\sum_{n_t} \sinh (r_{n_t} \Gamma) \cosh (r_{n_t} \Gamma) \bigg|_{\Gamma \ll 1} &\simeq \sum_{n_t} r_{n_t} \Gamma = \Gamma \sqrt{\frac{\Omega_m}{\Omega_p}} \propto \langle \hat{n} \rangle^{1/2},
\end{align}
The excitation probability thus becomes proportional to the photon flux. 
With Eq.~(\ref{eq.photon-flux_single-spatial-mode}), 
we obtain the cross section
\begin{align}
R^{TPA, corr}_{low-gain, single-mode} &= N_{mol} \frac{\sigma^{(2)} }{ A_p T_e } \frac{ \gamma_{fg} }{ \Omega_p } \phi,
\end{align}
where $T_e = 2\pi / \Omega_m$ is the entanglement time~(\ref{eq.T_e}), and $A_p$ the transverse beam area. 
This coincides with Eq.~(\ref{eq.sigma_e}) of the previous section in the limit of a narrow resonance, with the entanglement area $A_e$~(\ref{eq.A_e}) replaced by the transverse beam area $A_p$. 
At high gain $\Gamma \gg 1$, we approximate
\begin{align}
\sum_{n_t} \sinh (r_{n_t} \Gamma) \cosh (r_{n_t} \Gamma) \bigg|_{\Gamma \gg 1} &\simeq \frac{1}{4} \sum_{n_t} e^{2 r_{n_t} \Gamma} \simeq \langle \hat{n} \rangle / 2.
\end{align}
This means, the excitation probability is proportional to the squared photon flux, and we obtain the TPA rate at high-gain 
\begin{align} \label{R^corr_high-gain}
R^{TPA, corr}_{high-gain, single-mode} &= N_{mol} \sigma^{(2)} \frac{ \gamma_{fg}}{2 \Omega_p } \frac{ 1 }{ T_{pulse} f_{rep} } \times \phi^2.
\end{align}
Here we encounter the product of the laser pulse duration $T_{pulse} = 2\pi / \Omega_p$ and the laser repetition rate $f_{rep}$. This factor is also encountered in TPA with ultrafast lasers, and enhances the pulsed TPA absorption cross section relative to the sample's cw cross section when $T_{pulse} f_{rep} \ll 1$.

\subsection{Uncorrelated contribution}
Just like in the above calculation of the correlated contribution, we find
\begin{widetext}
\begin{align} 
P_f^{unc} (\vec{r}_{mol}) 
&= 2 \left( \frac{\omega_p / 2}{\hbar \epsilon_0 n_0 c} \right)^2 \sum_{e, e'} \frac{\mu^\ast_{ge} \mu^\ast_{ef} \mu_{ge'} \mu_{e'f}}{ (\omega_{eg} - \omega_0) (\omega_0 - \omega_{fe'}) } \tilde{h}^4 (\vec{r}_{mol}) \sum_{n_t, n_t'} \sinh^2 (r_{n_t} \Gamma)\sinh^2 (r_{n_t'} \Gamma) \notag\\
&\times \int dw \frac{1}{\pi} \frac{ w_{fg}}{w_{fg}^2 + (\frac{\omega_{fg} - \omega_p}{ \sqrt{2\Omega_m\Omega_p} } + w)^2} \left( \sqrt{\frac{n!}{m!}} w^{m-n}  L^{(m-n)}_n \left(w^2 \right) e^{- w^2 / 2} \right)^2, \label{eq.P_unc-exact}
\end{align}
\end{widetext}
where $m = \text{max} (n_t, n_t')$, and $n = \text{min} (n_t, n_t')$. 

\begin{figure*}
\includegraphics[width=\textwidth]{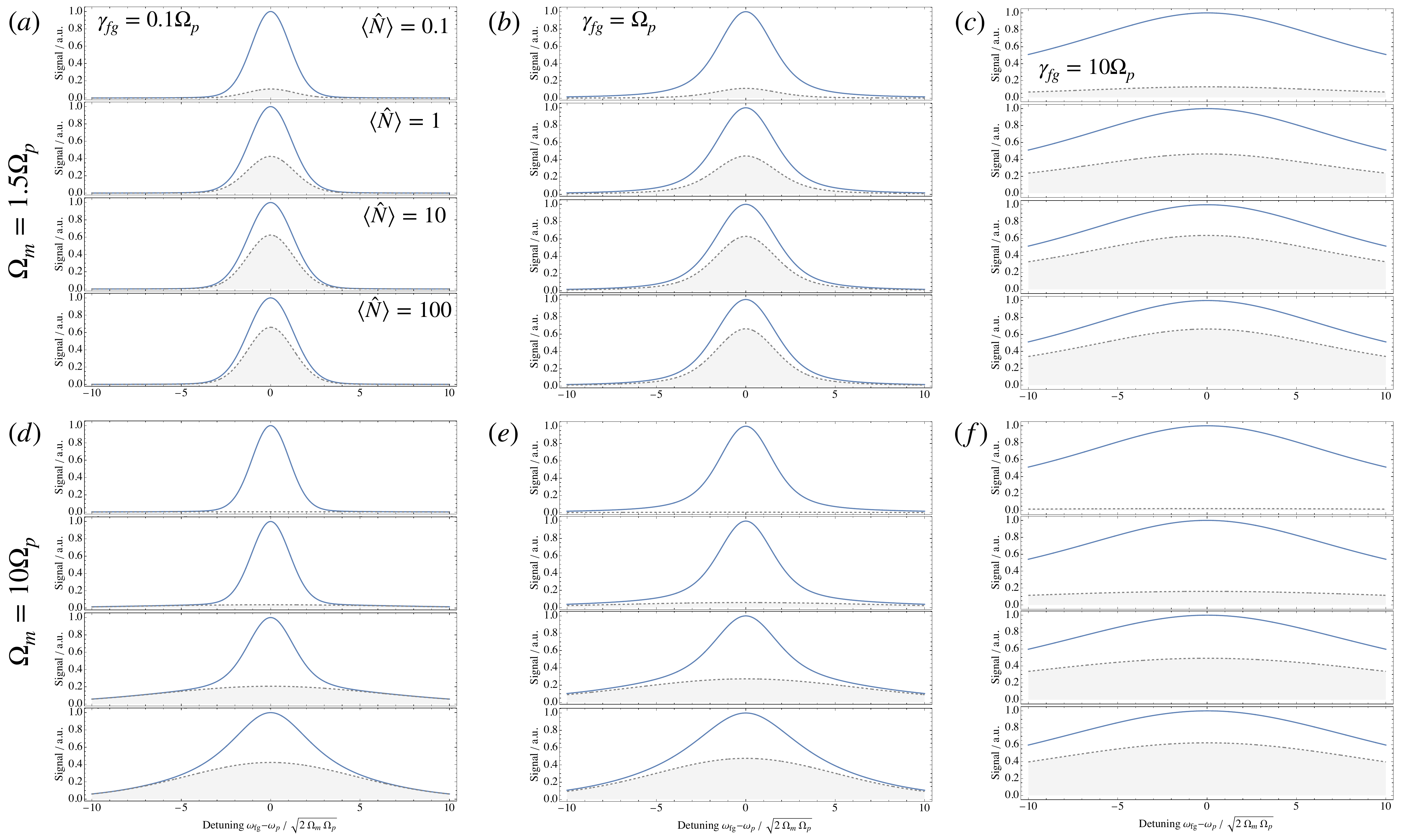}
\centering
\caption{
\textbf{Single spatial mode }
The ETPA resonance according to Eqs.~(\ref{eq.P_f^corr1}) and (\ref{eq.P_unc-exact}) is plotted (solid blue line) at different mean photon numbers, as indicated, and (a) $\gamma_{fg} = 0.1 \Omega_p$ and $\Omega_m = 1.5 \Omega_p$, (b) $\gamma_{fg} = \Omega_p$ and $\Omega_m = 1.5 \Omega_p$, (c) $\gamma_{fg} = 10 \Omega_p$ and $\Omega_m = 1.5 \Omega_p$, (d) $\gamma_{fg} = 0.1 \Omega_p$ and $\Omega_m = 10 \Omega_p$, (e) $\gamma_{fg} = \Omega_p$ and $\Omega_m = 10 \Omega_p$, and (f) $\gamma_{fg} = 10 \Omega_p$ and $\Omega_m = 10 \Omega_p$. 
The gray dashed line indicates the uncorrelated contribution~(\ref{eq.P_unc-exact}) to the full ETPA signal. Each plot is normalized to its maximal value (i.e. on resonance). 
}
\label{fig.resonance}
\end{figure*}

\paragraph{Narrow resonance}
We again replace the Lorentzian resonance by a delta-function~(\ref{eq.delta-approximation}). 
On resonance, when $\omega_{fg} = \omega_p$, Eq.~(\ref{eq.P_unc-exact}) then simplifies to
\begin{align} \label{eq.P_f-delta2}
P_f^{unc} (\vec{r}_{mol}) &= \left( \frac{\omega_p / 2}{\hbar \epsilon_0 n_0 c} \right)^2 \sum_{e, e'} \frac{\mu^\ast_{ge} \mu^\ast_{ef} \mu_{ge'} \mu_{e'f}}{ (\omega_{eg} - \omega_0) (\omega_0 - \omega_{fe'}) } \tilde{h}^4 (\vec{r}_{mol}) \notag \\
&\times \sum_{n_t} \sinh^4 (r_{n_t} \Gamma).
\end{align} 
To extract the TPA cross section from this expression, we wish to express the corresponding TPA signal as $\delta_r \phi^2$, where $\phi^2 \propto (\sum_n \sinh^2 (r_n \Gamma))^2$, see Eq.~(\ref{eq.photon-flux_single-spatial-mode}). We thus need to analyze the term $\sum_n \sinh^4 (r_n \Gamma)$. 
Given the Schmidt number $K$ of the PDC light, we expect that in the limit of many Schmidt modes, we have $\sim K$ terms which contribute to this expression. In contrast, the squared mean photon number should contain $\sim K^2$ terms. Thus, in a highly entangled light field, when the parameter $\zeta_t$ in Eq.~(\ref{eq.zeta_t}) is close to one, we expect that 
\begin{align}
\sum_{n_t} \sinh^4 (r_{n_t} \Gamma) \sim \frac{1}{K} \left( \sum_{n_t} \sinh^2 (r_{n_t} \Gamma) \right)^2.
\end{align}
This should be true at least for sufficiently small photon numbers, when many Schmidt modes contribute to the signal. 
In this regime, we obtain 
\begin{align} \label{R^unc_high-gain}
R^{TPA, unc}_{high-gain, single-mode} &= N_{mol} \sigma^{(2)} \frac{\gamma_{fg}}{ \Omega_m } \frac{ 1 }{ T_{pulse} f_{rep} } \times \phi^2.
\end{align}
Hence, in contrast to the correlated contribution~(\ref{R^corr_high-gain}), where the ratio $\gamma_{fg} / \Omega_p$ appears, here it is the ratio $\gamma_{fg} / \Omega_m$, i.e. the ratio between the molecular resonance and the bandwidth $\Omega_m$ of the individual photons. As $\Omega_m \gg \Omega_p$ in a highly entangled beam, this reduces the uncorrelated contribution to the cross section considerably.  

This behaviour again changes in a very high gain regime, when $e^{r_0 \Gamma} \gg e^{r_1 \Gamma}$. In this regime, the largest Schmidt mode becomes dominant, and the ETPA signal approaches that of a single-mode squeezed state in this mode. Then the spectral correlations lose their importance, and the uncorrelated TPA rate becomes exactly twice the correlated contribution~(\ref{R^corr_high-gain}),
\begin{align} \label{eq.R^TPA_very-high-gain}
R^{TPA, unc}_{very high-gain, single-mode} &=2 R^{TPA, corr}_{high-gain, single-mode}.
\end{align}

\subsection{Results}

In the final part of this Section, we analyze the results obtained thus far, and discuss physical implications. 

\begin{figure}
\includegraphics[width=0.45\textwidth, trim={60 0 0 0}, clip]{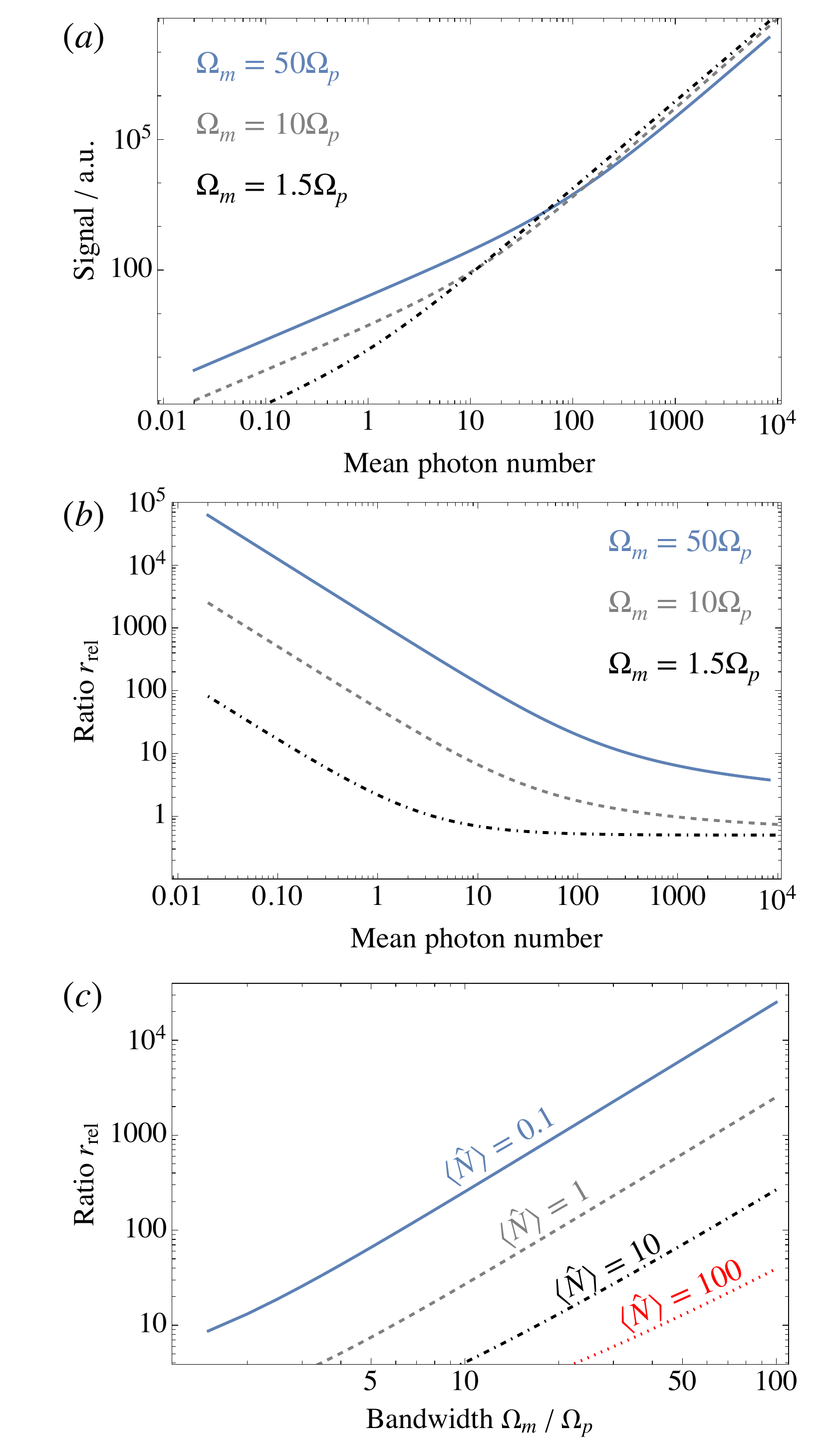}
\centering
\caption{
\textbf{Single spatial mode }
(a) ETPA Signal strength~(\ref{eq.P_f-delta}) and (\ref{eq.P_f-delta2}) vs. the mean photon number with bandwidths $\Omega_m = 50 \Omega_p$ (blue, solid line), $10 \Omega_p$ (gray, dashed), and $1.5 \Omega_p$ (black, dot-dashed). Here, a narrow spectral resonance is assumed, see Eq.~(\ref{eq.delta-approximation}). 
(b) Ratio between correlated and uncorrelated contributions, Eq.~(\ref{eq.r_rel}), for the same parameters an in (a). 
(c) Ratio~(\ref{eq.r_rel}) as a function of the bandwidth at fixed mean photon number $\langle \hat{N} \rangle = 0.1$ (blue, solid), $1$ (gray dashed), $10$ (black, dot-dashed), and $100$ (red, dotted). 
}
\label{fig.crossover}
\end{figure}

\textit{Spectral Resonance--}
In Fig.~\ref{fig.resonance}, we solve Eqs.~(\ref{eq.P_f^corr1}) and (\ref{eq.P_unc-exact}) numerically to show the change of the ETPA resonance as function of the broadening $\gamma_{fg}$, the bandwidth $\Omega_m$ (or, equivalently, the amount of entanglement), and the mean photon number $\langle \hat{N} \rangle$ per entangled pulse. The simulations show the expected behaviour, where at small photon numbers $\langle \hat{n} \rangle = 0.1$, the signal is dominated by the correlated contribution~(\ref{eq.P_f^corr1}). As the mean photon number increases, the incoherent part~(\ref{eq.P_unc-exact}) becomes more dominant. However, the crossover photon flux, where the uncorrelated events become more likely than the correlated ones depends strongly on the broadening of the final state resonance and the amount of entanglement (the Schmidt number) of the light field. 
For instance, in Fig.~\ref{fig.resonance}(c), where the resonance is very broad and the light field is only very weakly entangled, the uncorrelated events account for $\sim 50$~\% of the signal at $\langle \hat{n} \rangle = 1$. Conversely, in Fig.~\ref{fig.resonance}(d), we simulate a narrow resonance and strong entanglement of the incident light. Even at $\langle \hat{n} \rangle = 100$, the correlated contribution still accounts for about $60$~\%  of the signal. 

\textit{Intensity dependence of narrow molecular resonance--}
The transition of a nonlinear optical signal from the separate photon pairs to a multi-photon state was first analyzed in Ref.~\onlinecite{Schlawin2013}. 
In the following, we explore this behaviour in the limit~(\ref{eq.delta-approximation}), were $\gamma_{fg} \ll \Omega_p$. 
The increase of the full ETPA signal on resonance is shown in Fig.~\ref{fig.crossover}(a) for three different bandwidths $\Omega_m$ ranging from $1.5\Omega_p$ (very weak entanglement) to $50 \Omega_p$ (very strong entanglement). At photon numbers below $\langle \hat{N} \rangle \lesssim 100$, the strong quantum correlations in the latter case afford a large advantage over the former one. Moreover, whereas the linear signal scaling gives way to a quadratic scaling at $\langle \hat{N} \rangle \simeq 1$ in the case of weak entanglement, it remains linear up to $\langle \hat{N} \rangle \simeq 100$ in the strongly entangled case. 
The crossover takes place when the mean photon number per Schmidt mode $n_t$ becomes of order one, i.e. when $\langle \hat{A}^\dagger_{n_t} \hat{A}_{n_t} \rangle \sim 1$. Using the Schmidt number $K$ as the effective number of modes, we expect this to be the case when $\langle \hat{N} \rangle / K \sim 1$. In the case of large bandwidths, $\Omega_m = 50 \Omega_p$, we have, using Eq.~(\ref{eq.Schmidt-number}), $K \simeq \Omega_m / (2\Omega_p)$ and thus expect the crossover to take place when $\langle \hat{N} \rangle \simeq 25$. This estimation is consistent with our observations, given that the Schmidt number only gives an estimate for the effective dimensionality of the entangled state. 
 
However, this extended linear regime is not necessarily an advantage in terms of absolute signal strength, as Fig.~\ref{fig.crossover}(a) also reveals: when we increase the mean photon number even further, $\langle \hat{N} \rangle > 100$, the weakly entangled signal with $\Omega_m = 1.5 \Omega_p$ becomes \textit{stronger} than the other two. This is entirely due to the larger strength of the uncorrelated contribution on resonance. Its bandwidth is narrower, and thus gives rise to a narrower resonance, too [compare Fig.~\ref{fig.resonance}(a)]. On balance, this can be enough to overpower the much larger correlated contributions of highly entangled states. 
It appears that at these very large photon numbers, a few-mode or single-mode entangled state is in fact preferable over a (highly entangled) multimode state - at least insofar as the total signal strength is concerned. 
But in many proposed applications of entangled photons for spectroscopy, one is instead interested, e.g., in exploiting the quantum correlations of the entangled pulses to control the excited states in a multilevel system~\cite{NatComm13, Raymer2013, Oka2011, Oka2011b, Svozilik2018, Svozilik2018b, Roberto2019, Fujihashi19, Ishizaki2020, Ishizaki2021}, and this control is enabled by the absorption of pairs of correlated photons. The present results thus demonstrate that such a control is feasible not only in a separate pair limit, where the overall signal strength is very low. It can also be achieved with pulses containing much larger photon numbers. However, these states may not be optimal for inducing the largest possible ETPA signal, and one has to accept the uncorrelated background signal. 

We next analyze in more detail the relative size of correlated and uncorrelated contributions,
\begin{align} \label{eq.r_rel}
r_{rel} \equiv \frac{ P_f^{corr} }{ P_f^{unc} },
\end{align}
where the two terms are calculated with Eqs.~(\ref{eq.P_f-delta}) and (\ref{eq.P_f-delta2}), respectively. It is shown in Fig.~\ref{fig.crossover}(b) as a function of the mean photon number. 
Starting from very large values when $\langle \hat{N} \rangle < 1$, the ratio decreases steadily with increasing photon number, until it saturates to $r_{rel} \rightarrow 1/2$, which is the expected limiting behaviour according to Eq.~(\ref{eq.R^TPA_very-high-gain}). This limit can be reached already at $\langle \hat{N} \rangle \simeq 10$ for a weakly entangled state, and can be $\langle \hat{N} \rangle > 10^4$ for strongly entangled ones. 
Conversely, at fixed photon number, the ratio increases quadratically with the bandwidth $\Omega_m$, as is shown in Fig.~\ref{fig.crossover}(c). 

\textit{Crossover from narrow to broad molecular resonance--}
\begin{figure}
\includegraphics[width=0.49\textwidth]{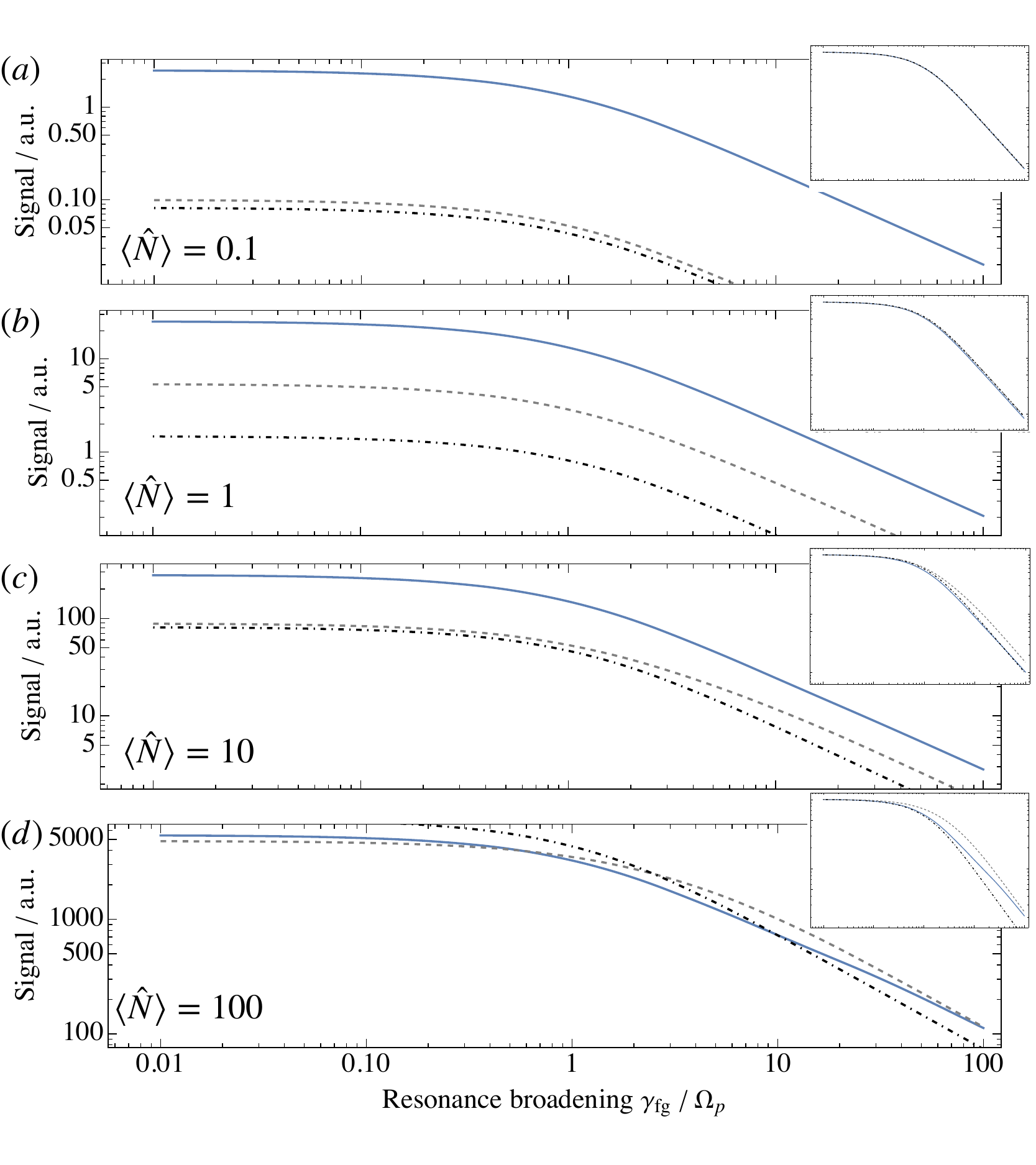}
\centering
\caption{
\textbf{Single spatial mode }
ETPA signal according to Eqs.~(\ref{eq.P_f^corr1}) and (\ref{eq.P_unc-exact}) is plotted vs. the broadening $\gamma_{fg}$ at fixed photon numbers (a) $\langle \hat{N} \rangle = 0.1$, (b) $\langle \hat{N} \rangle = 1$, (c) $\langle \hat{N} \rangle = 10$, and (d) $\langle \hat{N} \rangle = 100$, and with bandwidths $\Omega_m = 50 \Omega_p$ (blue, solid line), $10 \Omega_p$ (gray, dashed), and $1.5 \Omega_p$ (black, dot-dashed). 
The small insets show the same plots, where each plot is normalized to its limiting value at $\gamma_{fg} \rightarrow 0$. 
}
\label{fig.narrow-to-broad}
\end{figure}
So far, our discussion was limited to a very narrow molecular resonance. The dependence of the ETPA signal on a finite resonance is investigated in Fig.~\ref{fig.narrow-to-broad}, where we solve Eqs.~(\ref{eq.P_f^corr1}) and (\ref{eq.P_unc-exact}) numerically to obtain $R^{TPA}$. 
For a very broad range of mean photon numbers $\langle \hat{N} \rangle$ and bandwidths $\Omega_m$, we find that the signals saturate to a constant value when $\gamma_{fg} \lesssim 0.1 \Omega_p$. We have checked that this limit coincides with the delta-function limits, Eqs.~(\ref{eq.P_f-delta}) and (\ref{eq.P_f-delta2}), which provide excellent approximations in this parameter regime. 
When the broadening becomes similar to or larger than the pump bandwidth $\Omega_p$, the signal starts to decay with increasing $\gamma_{fg}$. 
If the mean photon number is small and the signal is dominated by the correlated contribution to $P_f$, this decease is independent of the photonic bandwidth $\Omega_m$. This can be seen in the insets of Figs.~\ref{fig.narrow-to-broad}(a) and (b), where the plots are normalized to their maximal value, and the three plots corresponding to different bandwidths fall on top of one another. 
In all cases, we have $R^{TPA} \propto 1 / \gamma_{fg}$. 
At larger photon numbers, when the incoherent contribution to $P_f$ becomes important as well, this universal scaling behaviour is lost. 
As observed already in Fig.~\ref{fig.crossover}(a), a state with narrower bandwidth (black, dot-dashed line) generates a larger ETPA signal than a strongly entangled state (blue solid line) when the resonance broadening is small. 
But when the broadening becomes larger, this changes once more, and e.g. in Fig.~\ref{fig.narrow-to-broad}(d), the strongly entangled state generates the largest ETPA signal, when $\gamma_{fg} \gtrsim 100 \Omega_p$ - albeit a much smaller absolute signal than in the case of a narrow resonance. 

\begin{figure*}
\includegraphics[width=0.75\textwidth, trim=10 280 100 0]{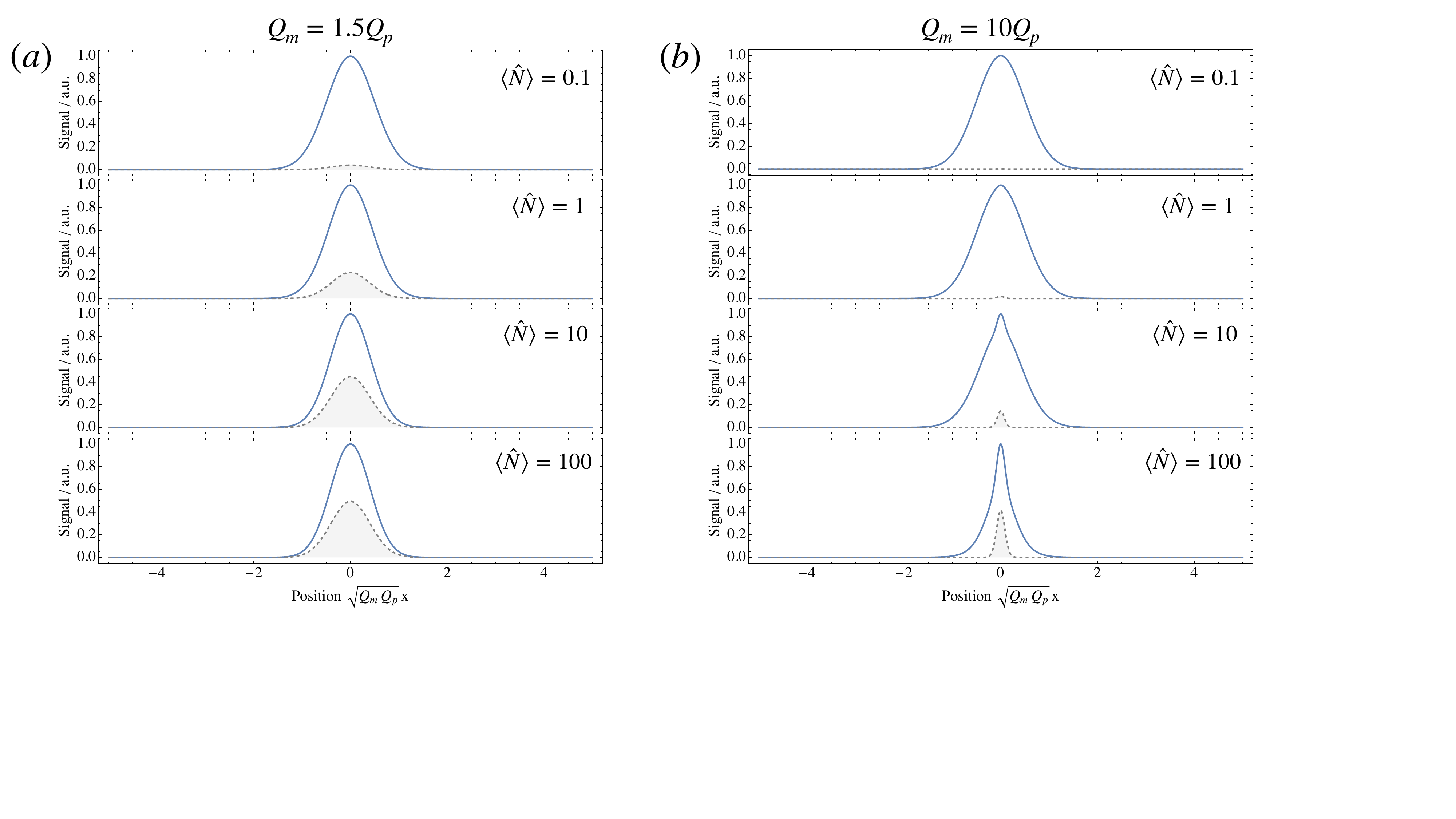}
\centering
\caption{
\textbf{Single spectral mode }
(a) ETPA Signal strength according to Eqs.~(\ref{eq.P_f^corr-spatial}) and (\ref{eq.P_f^unc-spatial}) is shown vs. one spatial direction $x$, with $y = 0$. 
The gray shaded area indicates the uncorrelated contribution~(\ref{eq.P_f^unc-spatial}) to the full ETPA signal.
(b) the same as (a), but with stronger spatial correlations with $Q_m = 10 Q_p$.
}
\label{fig.spatial-resonance}
\end{figure*}

\section{Single spectral mode}
\label{sec.single-spectral-mode}
We next turn to the analysis of the third limiting case, which we discuss in this manuscript, where only a single spectral mode must be considered. Such a state could be prepared, for instance, by spectral filtering of the PDC light (at the cost of reducing the photon flux). 
This situation is easier to deal with than that the previous Section~\ref{sec.single-spatial-mode}. It basically amounts to analysing the probability of localising two photons simultaneously at the position of the absorber. A molecular response function is not involved (or rather, it only gives the same contribution to correlated and uncorrelated terms in the final signal). 
In the following, we use the following single spectral mode function
\begin{align}
\tilde{h} (\omega) &= (\pi \Omega_p^2)^{- 1/4} \exp \left[ - \frac{(\omega- \omega_p / 2)}{2\Omega_p^2} \right] 
\end{align} 
which we obtain from Eq.~(\ref{eq.F_spec}) by setting $\Omega_m = \Omega_p$. 
Adapting Eq.~(\ref{eq.P_f-final}) for this situation, we obtain
\begin{align} \label{eq.P_f^corr-spatial}
&P_f^{corr} = \sigma^{(2)} 2\gamma_{fg} \text{spec} (\omega_p - \omega_{fg}, \Omega_p, \gamma_{fg}) \notag \\
&\times \!\!\!\!\!\!\!\!\! \sum_{n_x, n_y,n'_x, n'_y} \!\!\!\!\!\! \sinh (r_{n_x, n_y} \Gamma) \cosh (r_{n_x, n_y} \Gamma) \sinh (r_{n'_x, n'_y} \Gamma) \cosh (r_{n'_x, n'_y} \Gamma)  \notag \\
&\times h_{n_x}^2 (\sqrt{Q_mQ_p}x) h_{n'_x}^2 (\sqrt{Q_mQ_p}x) h_{n_y}^2 (\sqrt{Q_mQ_p}y) h_{n'_y}^2 (\sqrt{Q_mQ_p}y).
\end{align}
The uncorrelated contribution likewise simplifies to
\begin{align} \label{eq.P_f^unc-spatial}
&P_f^{unc} = \sigma^{(2)} 4 \gamma_{fg} \text{spec} (\omega_p - \omega_{fg}, \Omega_p, \gamma_{fg}) \notag \\
&\times  \sum_{n_x, n_y} \sum_{n'_x, n'_y} (-1)^{n_x + n_y+n'_x+n'_y} \sinh^2 (r_{n_x, n_y} \Gamma ) \sinh^2 (r_{n'_x, n'_y} \Gamma ) \notag \\
&\times h_{n_x}^2 (\sqrt{Q_mQ_p}x) h_{n'_x}^2 (\sqrt{Q_mQ_p}x) h_{n_y}^2 (\sqrt{Q_mQ_p}y) h_{n'_y}^2 (\sqrt{Q_mQ_p}y). 
\end{align}
In these expressions, we have combined the spectral integrals into the function
\begin{align}
&\text{spec} (\omega_p - \omega_{fg}, \Omega_p, \gamma_{fg}) \notag\\ 
&= \int d\omega_1 \int d\omega_2 \int d\omega_{sum} \frac{1}{\pi} \frac{\gamma_{fg} }{ \gamma_{fg}^2 + (\omega_{fg} - \omega_{sum})^2 } \notag \\ 
&\quad\times \tilde{h} (\omega_{sum} - \omega_1) \tilde{h} (\omega_1) \tilde{h} (\omega_{sum} - \omega_2) \tilde{h} (\omega_2) \notag \\
&= \Re \tilde{w} (\omega_p - \omega_{fg} + i \gamma_{fg}),
\end{align}
which describes the overlap between the spectral mode function and the molecular response. In the present case, where we assumed Gaussian spectral modes, it evaluate to the real part of the Faddeeva function $\tilde{w}$, as the response is given by the convolution of a Gaussian and a Lorentzian, i.e. by a Voigt profile. 
The only difference between the two terms in Eqs.~(\ref{eq.P_f^corr-spatial}) and (\ref{eq.P_f^unc-spatial}) - apart from a factor 2 - is the scaling behaviour $\sim \sinh^2 \cosh^2$ vs $\sim \sinh^4$ and the factor $(-1)^{n_x + n_y+n'_x+n'_y}$, which reduces the uncorrelated contribution, respectively. This contrasts with the previous section, where correlated and uncorrelated contributions were convoluted differently with the molecular response function, even in the limit of a a very narrow resonance~(\ref{eq.delta-approximation}). 

The spatial dependence of the of the signal according to Eqs.~(\ref{eq.P_f^corr-spatial}) and (\ref{eq.P_f^unc-spatial}) is shown in Fig.~\ref{fig.spatial-resonance} for different mean photon numbers and a weakly entangled state with $Q_m = 1.5 Q_p$, and a entangled state with $Q_m = 10 Q_p$.  The uncorrelated contribution is shown as a dashed, gray line. In the strongly entangled case, the latter is much more narrowly peaked, signifying that ETPA is possible only in a very narrow area with the largest photon density. 

\begin{figure}
\includegraphics[width=0.4\textwidth]{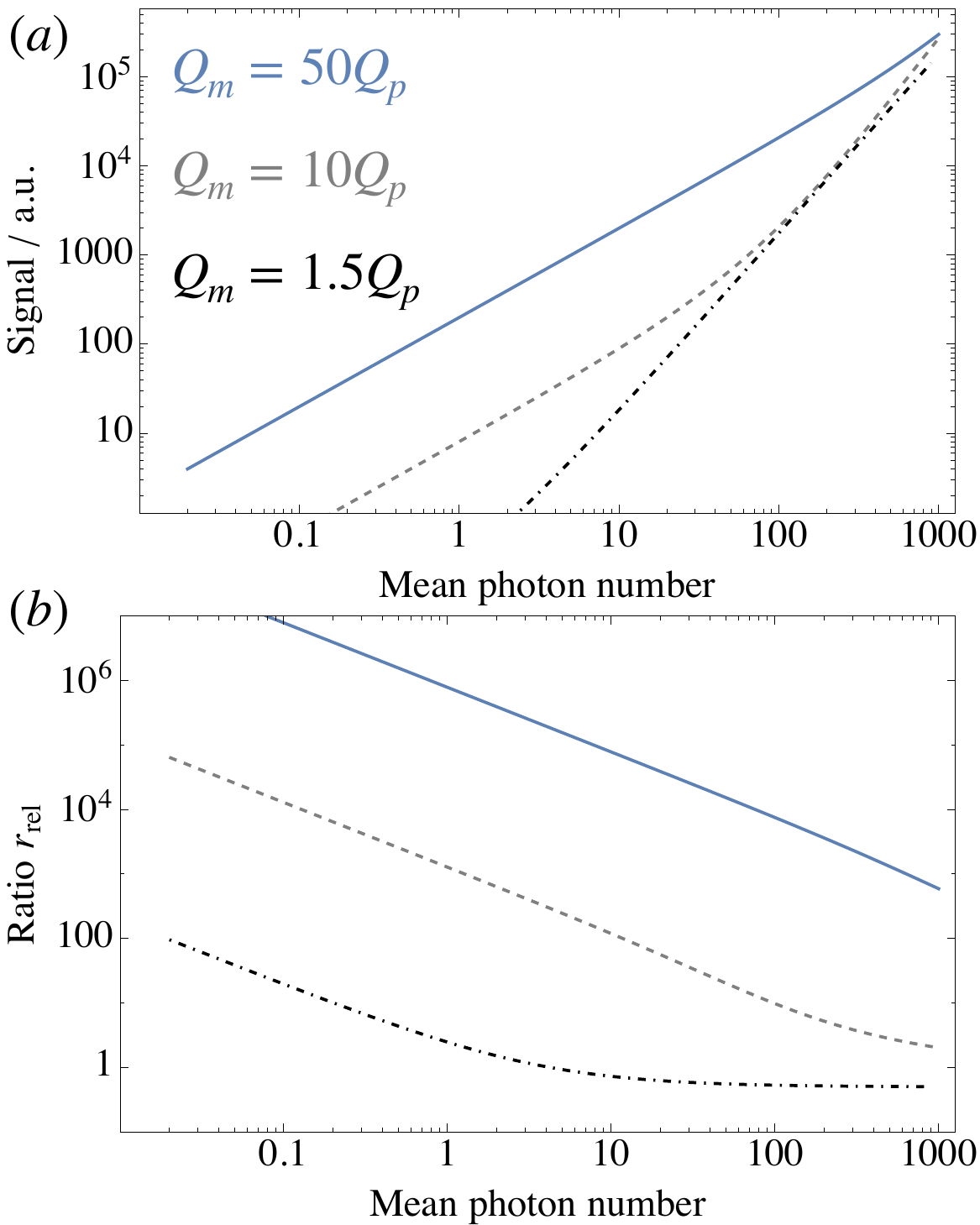}
\centering
\caption{
\textbf{Single spectral mode }
(a) ETPA signal strength according to Eqs.~(\ref{eq.R_TPA-def}), (\ref{eq.P_f^corr-spatial}) and (\ref{eq.P_f^unc-spatial}) is shown vs. the mean photon number. 
(b) The ratio $r_{rel}$, Eq.~(\ref{eq.r_rel}), extracted from the signals in (a). 
}
\label{fig.spatial-signal}
\end{figure}
In Fig.~\ref{fig.spatial-signal}, we basically repeat the analysis of the previous section, and analyze the scaling behaviour of the ETPA signal~(\ref{eq.R_TPA-def}), which we obtain from integrating Eqs.~(\ref{eq.P_f^corr-spatial}) and (\ref{eq.P_f^unc-spatial}) over space. 
As before, we show both weakly entangled states (with $Q_m = 1.5 Q_p$) and very strongly entangled ones ($Q_m = 50 Q_p$). 
Here we find that, in contrast to Fig.~\ref{fig.crossover}, the strongly correlated signals always remain larger than weakly entangled ones. The reason is straightforward: the spectral correlations are convoluted with the molecular response, which gives rise to the intricate interplay we had explored in the previous section. 
The spatial correlations, on the other hand, manifest themselves a distinct scaling behaviour of Eqs.~(\ref{eq.P_f^corr-spatial}) and (\ref{eq.P_f^unc-spatial}), which are otherwise identical. Consequently, the advantage due to spatial quantum correlations become smaller in relative terms as the mean photon number increases, but it never goes away and may persist even to photonic states with macroscopic photon numbers~\cite{Spasibko17, Cutipa2021}.

\section{Conclusions}

In this manuscript, we have investigated entangled two-photon absorption cross sections of pulsed entangled beams from the low- to the high-gain regime of parametric downconversion. 
We have first presented a derivation of Fei's seminal formula for the ETPA cross section~(\ref{eq.ETPA-cross-section_Fei}) starting from a microscopic model of light. 
Eq.~(\ref{eq.ETPA-cross-section_Fei}) relies on a factorization of the entangled photon wavefunction into spatial and spectral components. When these degrees of freedom are quantum correlated \cite{Brambilla2004, Gatti2009}, such that they cannot be factorized, the above formula does not necessarily apply and new effects relying on the joint spatio-temporal entanglement might be expected. 

Our analysis shows that the formula~(\ref{eq.ETPA-cross-section_Fei}) proposed by Fei et al. applies only in the limit of a very narrow (compared to the pump laser bandwidth) molecular resonance. A more realistic treatment, which accounts for a finite broadening, gives rise to the result~(\ref{eq.sigma_e}). The finite resonance broadening leads to a reduction of the cross section when the experiment is conducted with continuous-wave entangled photon pairs.  
This finding could explain some of the unsuccessful attempts at detecting ETPA with continuous-wave entangled photon pairs. 
Moreover, we have clarified the microscopic origin of the entanglement time and the entanglement area, which appear in Eq.~(\ref{eq.ETPA-cross-section_Fei}). These quantities are directly connected to the amount of entanglement in the photonic state only in certain limits (i.e. when $\Omega_m \gg \Omega_p$ and $Q_m \gg Q_p$). Still, even in situations where this is not the case ( for instance, when both $\Omega_m$ and $\Omega_p$ are large), small entanglement times and the entanglement areas are always beneficial in enhancing the ETPA cross section. 
In such a situation, we may still have \textit{quantum} enhancement (due to the linear scaling brought about by the two-photon nature of the state), but it is not due to quantum entanglement. 
As both the entanglement time and the entanglement area are controlled by properties of the nonlinear crystal where the entangled photons are generated, these findings should be reflected in the design of novel quantum light sources for spectroscopy~\cite{Moretti2023}. 

We have further investigated the crossover from linear to quadratic regime of ETPA. We have seen that ETPA can benefit from the same enhancement as TPA with ultrafast lasers. We have further shown that anticipated crossover behaviour according to Eq.~(\ref{eq.ETPA-rate_Fei}) is incorrect, as the probability for the absorption of correlated photon pairs also increases quadratically with the photon number at sufficiently large photon fluxes. Thus, the cross section in the quadratic regime is composed of two contributions, and quantum enhancement effects can persist to very large photon numbers. 
They are lost only when the multimode nature of the light field becomes negligible. We point out, however, that in this very high-gain regime, corrections to the simple PDC model we employed in this manuscript may become substantial and will require a numerically more involved treatment of the light source~\cite{Christ2013}. 
In this discussion, we have considered degenerate PDC. This is in fact the worst case scenario, where the uncorrelated contributions can als drive TPA resonantly. In non-degenerate PDC, this background would be suppressed. 

In this manuscript, we have only treated the impact of spatial and spectral correlations separately. As one can see from the basic equations we derived here, their interplay may become highly nontrivial and give rise to interesting new effects in the high gain regime, which could be very appealing, e.g., for applications in nonlinear quantum imaging~\cite{Moreau2019, Ma2021, Bowen2023} or spatially resolved spectroscopy. This will be an interesting direction for future research. 
Overall, our work highlights the interesting effects one can expect from nonlinear light-matter interactions of high-gain PDC light~\cite{Gilaberte2019, Spasibko17, Cutipa2021}. 

\begin{acknowledgments}
I would like to thank Dr. Shahram Panahiyan for his helpful feedback on this manuscript, and acknowledge support from the Cluster of Excellence 'Advanced Imaging of Matter' of the Deutsche Forschungsgemeinschaft (DFG) - EXC 2056 - project ID 390715994.
\end{acknowledgments}

%

\bibliography{bibliography_PDC}

\end{document}